\documentclass[nopreprintline,5p,twocolumn,authoryear]{arxiv_elsarticle}

\usepackage{multirow} 

\usepackage{hyperref}
\usepackage{makecell}

\usepackage[font=footnotesize,caption=false]{subfig}

\usepackage[cmex10]{amsmath}						
\usepackage{amsfonts}						
\usepackage{amssymb}
\DeclareMathOperator*{\argmax}{arg\,max}

\usepackage[latin1]{inputenc} 

\journal{Journal of Network and Computer Applications}

\begin{document}

\begin{frontmatter}

\title{A Survey on QoE-oriented Wireless Resources Scheduling}

\author{Ivo~Sousa$^\ast$} 
\ead{ivo.sousa@lx.it.pt}
\author{Maria~Paula~Queluz} 
\author{António~Rodrigues} 

\cortext[cor1]{Corresponding author}

\address{Instituto de Telecomunicações, Instituto Superior Técnico, University of Lisbon, 1049-001 Lisbon, Portugal}

\begin{abstract}

Future wireless systems are expected to provide a wide range of services to more and more users. Advanced scheduling strategies thus arise not only to perform efficient radio resource management, but also to provide fairness among the users. On the other hand, the users' perceived quality, i.e., Quality of Experience (QoE), is becoming one of the main drivers within the schedulers design. 
In this context, this paper starts by providing a comprehension of what is QoE and an overview of the evolution of wireless scheduling techniques. Afterwards, a survey on the most recent QoE-based scheduling strategies for wireless systems is presented, highlighting the application/service of the different approaches reported in the literature, as well as the parameters that were taken into account for QoE optimization. Therefore, this paper aims at helping readers interested in learning the basic concepts of QoE-oriented wireless resources scheduling, as well as getting in touch with its current research frontier.

\end{abstract}


\end{frontmatter}


\section{Introduction}
\label{section:intro}

Wireless resources scheduling comprises the allocation of physical radio resources among users and the determination of the users' serving order (also known as prioritization). The goal is to fulfill some service requirements such as fairness (including avoiding greedy users, where one user consumes all or almost all system resources) or congestion, along with other constrains like delay or packet loss rate.

Compared to wired networks, wireless channels have time-varying behaviors, hence more complex scheduling schemes are required for the latter. However, since the scheduling process allows to save resources, wireless schedulers play a crucial role in efficient management of scarce radio resources.

In order to improve the service level, wireless systems have adopted in the past years schedulers that provide Quality of Service (QoS), i.e., the network's capability to guarantee a certain level of performance to a data flow. Since QoS is usually evaluated in terms of delay, packet loss rate, jitter or throughput, QoS can be regarded as a service quality characterization that is network-centric.

Despite the popularity of QoS-oriented schedulers design, the end-users --- humans --- have the decisive judgment about the received service quality. In literature, some pioneering authors \citep{bib:4_of_16_of_001,bib:pioneer_mos,bib:gunther,bib:7_of_14_of_001} showed that the application of a subjective-based approach may lead to significant improvements on user perceived quality, i.e., Quality of Experience (QoE), compared to network-centric approaches, 
such as maximization of the system throughput (i.e., the sum of the data rates that are delivered to all terminals). 
Hence, a shift from QoS- to QoE-oriented mechanisms design has been observed in recent years.

QoE is a concept that tries to cover everything that a user experiences when dealing with multimedia services and systems \citep{bib:qualinet}; it takes into account not only the usability of a multimedia service or system, but also the information content.
Consequently, QoE can be regarded as a user-centric characterization of the service quality.

As the number of dimensions involved in the users' subjective evaluation is immense, QoE-based techniques are becoming progressively more complex and sophisticated than the previous QoS-oriented algorithms. Schedulers that make use of QoE features consequently try to directly reflect the subjective experiences of the users, resulting in their resource allocation and prioritization techniques to be more efficient in terms of satisfying the users than the schedulers that adopt conventional metrics. This efficiency can be achieved by avoiding wasting resources in situations where there is a small or even no impact on the user experience. Therefore, QoE-oriented wireless resources schedulers aim to fulfill the mobile system users expectations: watch/listen what I want, anywhere, anytime.

\begin{table*} [!tb] 
\footnotesize
\renewcommand{\arraystretch}{1.3}
\centering
\caption{Related existing surveys.}
\vspace{8pt}

\begin{tabular}{| m{0.1275\linewidth} | m{0.28\linewidth} | m{0.46\linewidth} |} 
\hline
\centering\arraybackslash\multirow{2}{*}{\textbf{Focus}} & \centering\arraybackslash\multirow{2}{*}{\textbf{References}} &  \centering\arraybackslash\multirow{2}{*}{\textbf{Remarks}} \\
& & \\
\hline\hline

QoE estimation for different types of services & \citet{bib:survey_video_2,bib:survey_video_1,bib:survey_video_qoe_1,bib:survey_video_qoe_2,bib:survey_parametric} & These works address QoE assessment models for several applications (e.g., video streaming, conversational voice, web browsing, file download), but disregard QoE provisioning methodologies. \\ 
\hline
QoE challenges with respect to mobile networks & \citet{bib:survey_challenges,bib:qoe_mobile,bib:challenges_mobile,bib:minimini_survey} & QoE monitoring and optimization issues are considered, although without fully addressing the scheduling of wireless resources. \\
\hline
Wireless resources scheduling & \citet{bib:survey_qos_wimax,bib:survey_qos_1,bib:survey_qos_oppo,bib:survey_qos_lte,bib:survey_qos_2,bib:survey_qos_3} & Only QoE-unaware strategies are contemplated. \\
\hline

%
%
%
%
%
\end{tabular}
\label{tab:surveys}
\end{table*}

Considering the existing literature, it was recognized a lack of a proper comprehensive guide regarding wireless schedulers design that take QoE into account (cf. Table~\ref{tab:surveys}): some works survey QoE models and assessment methods for a variety of services, but they do not consider QoE provisioning algorithms; other works focus on mobile networks and provide insights on QoE management issues, yet they do not perform an in-depth study of wireless resources schedulers; finally, some surveys addressed precisely the scheduling of wireless resources, but no QoE-aware procedures were reviewed.
Accordingly, this survey paper aims at filling this gap by giving an extensive overview of the key facets of QoE-oriented wireless resources scheduling.
Its main contributions are:
\begin{itemize}
 \item A taxonomy and classification of approaches for QoE-oriented resource scheduling in wireless networks;
 \item A survey of existing work, including the classification according to the aforementioned taxonomy;
 \item A brief discussion of each approach, giving the readers an idea about which parts of existing literature might be of interest to their requirements.
\end{itemize}
Therefore, this survey serves as a reference for those who want to implement QoE-aware wireless resource schedulers and also aims to be a valuable contribution for those who want to perform research within this topic.

\begin{figure}[!tb]
\begin{center}
\includegraphics[ width=1.0\columnwidth , trim= 0 0 0 0 ]{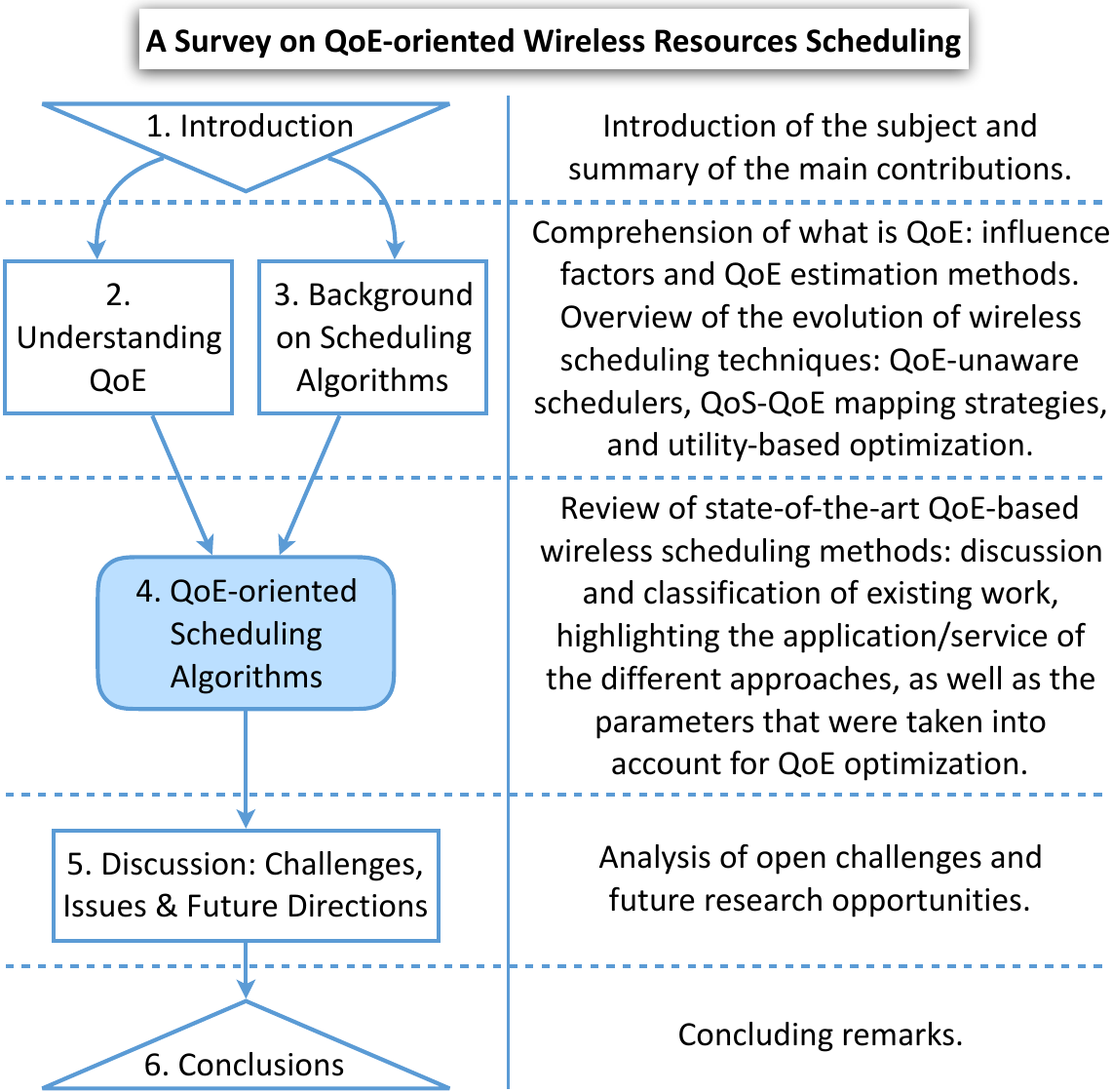}
\end{center}
\caption{Structure of this paper.}
\label{fig:structure}
\end{figure}

This paper is organized as follows (cf. Fig.~\ref{fig:structure}).
The first two sections that follow this introduction provide a contextualization regarding what is QoE and how traditional schedulers work: in Section~\ref{section:about_QoE}, the factors that influence the QoE in multimedia services over communication systems are presented, along with some QoE estimation methods; Section~\ref{section:back} illustrates some scheduling algorithms, ranging from the simplest ones to QoS-aware approaches, followed by the introduction of QoS-QoE mapping strategies and utility-based optimization. 
Section~\ref{section:QoE_algs} provides the main contribution of this survey, namely the presentation of recent research directions regarding QoE-oriented wireless resources scheduling --- state-of-the-art QoE-aware scheduling methods are discussed and classified based on the adjustments required, at the end-user devices, in order to implement the different scheduling strategies on wireless systems. 
Some of the important open challenges and future research opportunities are discussed in Section~\ref{section:disc_future}.
Finally, Section~\ref{section:conclusions} concludes the paper.


\section{Understanding QoE}
\label{section:about_QoE}

The Qualinet white paper on definitions of QoE states that ``QoE is the degree of delight or annoyance of the user of an application or service. It results from the fulfillment of his or her expectations with respect to the utility and/or enjoyment of the application or service in the light of the user's personality and current state'' \citep{bib:qualinet}.

\begin{figure*}[!tb]
\begin{center}
\includegraphics[ width=0.7\textwidth , trim= 0 0 0 0 ]{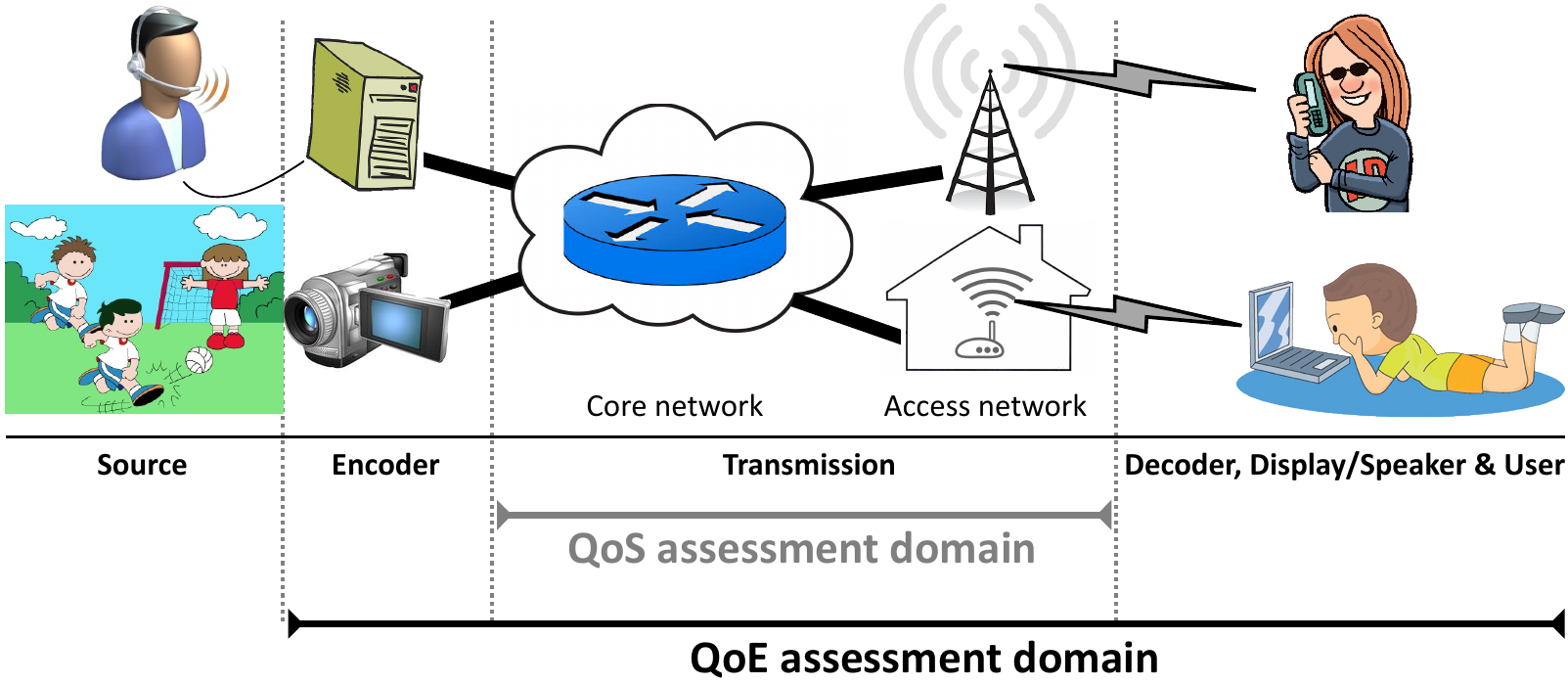}
\end{center}
\caption{Quality assessment domains in multimedia content delivery.}
\label{fig:QoS_QoE}
\end{figure*}

As far as communication systems are concerned, QoE can be affected by factors such as multimedia content, application, service, end-user device, network and context of use. 
For instance, \citet{bib:visual_contribution} showed that if supplementary visual observation of the speaker's facial and lip movements are also utilized besides the oral speech, higher levels of noise interference can be tolerated by humans than if no visual factors were taken into account. Hence, QoE assessment operations are performed in a broader domain when compared to QoS measurements  --- cf. Fig.~\ref{fig:QoS_QoE}.

The following subsections provide some details about the factors that influence the user experience, as well as some QoE estimation methods regarding multimedia services over communication systems.


\subsection{Factors Influencing QoE}
\label{subsection:qoe_facts}

According to the Qualinet white paper \citep{bib:qualinet}, an influence factor is defined as ``any characteristic of a user, system, service, application, or context whose actual state or setting may have influence on the QoE for the user''. Factors influencing QoE may be categorized as human, system, and context factors.


\subsubsection{Human factors} The characteristics of the users such as gender, age, and visual and auditory acuity are examples of human physical factors that may impact the users' perceived quality \citep{bib:human_geral}. On the other hand, more variant factors such as motivation, attention level, or users' mood, i.e., emotional factors, also play an important role when addressing the QoE influence factors \citep{bib:human_mood}. Moreover, even educational background, occupation, and nationality will affect the QoE \citep{bib:human_cultural}. In short, human factors that influence the perceived quality are complex and strongly interrelated, and their assessment should also take into consideration the time-dynamic perception of a service (i.e., the memory effect), where previous experiences also have influence on the current QoE \citep{bib:human_memory}.


\subsubsection{System factors} The technology employed for multimedia content transmission may introduce distortions or impairments in the content, which may affect the users' QoE. 
First, the original data need to be compressed, so that the multimedia content can be transmitted through a capacity-limited network. 
This encoding process, which incorporates many technical decisions such as the chosen bitrate (constant or variable), video frame rate or spatial resolution, may be lossless or lossy, meaning that the latter may lead to a quality degradation \citep{bib:system_frame_rate}. In addition, the transmission network may greatly affect the multimedia quality, namely due to major factors like packet loss, delay, and jitter \citep{bib:system_packet_delay}. Even with the adoption of a buffer at the receiver side, these network factors may cause the need for rebuffering, a streaming state activated when the playback buffer becomes empty and that leads to a playout stall, which is usually very annoying for the users \citep{bib:toon,bib:system_stall}. Considering non-streaming services, the task completion (e.g., the non completion of a download) or the excessive amount of time a service takes to download or to upload are QoE degrading situations \citep{bib:system_download}, which are also caused by packet delay or data flow rate reduction. Other system factors that may affect the perceived quality are the type of device used at the users' side (e.g., the screen resolution, user interface capabilities, audio loudness, computation power, or battery lifetime) and some system specifications (e.g., interoperability, personalization, security, or privacy) \citep{bib:system_battery} --- the reader is suggested to refer to \citep{bib:survey_challenges,bib:qoe_mobile,bib:challenges_mobile,bib:minimini_survey} for more examples and details on QoE challenges concerning mobile networks.

\subsubsection{Context factors} Apart from the two aforementioned group of factors, there are external factors that influence the users' QoE by affecting the surrounding environment \citep{bib:context_environment}. These context factors include temporal aspects, such as time of the day or day of the week (e.g., a better experience may be obtained when users are more relaxed, like during evenings or weekends), duration of the content and its popularity (e.g., users usually tolerate more distortion when they are watching popular videos), and service type, i.e., if it is live streaming or not (where users may have different quality expectations). The economic context can be also incorporated in this category of factors influencing QoE \citep{bib:context_business}, namely subscription type, costs, and brand of the system/service (including the availability of other service providers). 

\subsection{QoE Estimation Methods}
\label{subsection:measurement_methods}
Measuring and ensuring good QoE in multimedia applications is very subjective in nature. Hence, one way to assess QoE is to perform subjective tests, which directly measure the perceived quality by inquiring persons about their opinion regarding the quality of the multimedia content that is being tested. The subjective test results can also be used to validate the objective assessment performance, which is another quality assessment methodology.

The Mean Opinion Score (MOS), which is standardized by ITU-T \citep{bib:ITU_MOS}, is the most widely adopted QoE measurement.
MOS is defined as a numeric value ranging from 1 to 5 (1\nobreakdash-Bad, 2\nobreakdash-Poor, 3\nobreakdash-Fair, 4\nobreakdash-Good, 5\nobreakdash-Excellent) and it corresponds to the arithmetic mean of individual ratings in a panel of users. This approach has some drawbacks, 
namely it is costly, time consuming, and does not allow real-time evaluations.
Moreover, some useful information may not be captured (e.g., if an impairment occurs at a certain moment but affects the overall QoE, this particular moment may not be detected).

Objective quality methods have been developed in order to obtain a reliable QoE prediction while avoiding the need to perform subjective tests. The approach is based on mathematical techniques that yield quantitative measures of the multimedia content or service quality. 
Within the objective methods, two types of approaches can be identified: parameter-based methods and signal-based methods \citep{bib:approaches_objectives_methods}. The former rely on network/application parameters, such as viewing time \citep{bib:view_time}, download ratio \citep{bib:download_ratio_2,bib:download_ratio} or QoS parameters (Section~\ref{subsection:QoS_QoE_map} gives some examples of QoS-QoE mapping strategies). 
On the other hand, signal-based methods are based on the analysis of the signal; in \emph{intrusive} methods, the analysis compares the received data with a reference, which can be the full original data (full reference methods) or some key features of it (reduced reference methods); \emph{non-intrusive} methods, also known as no-reference methods, do not require access to the original multimedia content, relying only on the received signal to assess its quality.
Nevertheless, some issues arise when performing objective assessment. 
Although intrusive methods are generally accurate, they are impracticable for monitoring live transmissions due to the need of the original multimedia content. Also, objective assessment may not reflect the perception of the users concerning the delivered service; for example, although some impairments may cause minor influence on the users' QoE, and therefore they could be disregarded, the same impairments may be detected and emphasized by the objective methods.

It is also important to mention that the majority of QoE estimation methods that have been proposed so far address the specific case of video quality evaluation, mainly due to the popularity achieved by video streaming services over the last years --- surveys on video quality estimation can be found in \citep{bib:survey_video_2,bib:survey_video_1}, which mostly focus on objective quality models, and in \citep{bib:survey_video_qoe_1,bib:survey_video_qoe_2}, where the former work addresses metrics and methodologies relevant to the traditional video delivery, whereas the latter work focus on measurement mechanisms that are used to evaluate the QoE for online video streaming; a survey on parametric QoE estimation for popular services, such as video-on-demand streaming, Voice over IP (VoIP), web browsing, Skype and file download services, can be found in \citep{bib:survey_parametric}.

As can be inferred from what was presented so far, QoE cannot be easily modeled and assessed due to the fact that its influence factors are very diverse and they may interrelate, as well as different users have different quality expectations. In order to attain an enhanced QoE evaluation, a combination of objective and subjective methods can be carried out \citep{bib:neural_network_qos,bib_mix_sub_obj}. 

\section{Background on Scheduling Algorithms}
\label{section:back}

\begin{figure}[!tb]
\begin{center}
\includegraphics[ width=0.7\columnwidth , trim= 0 0 0 0 ]{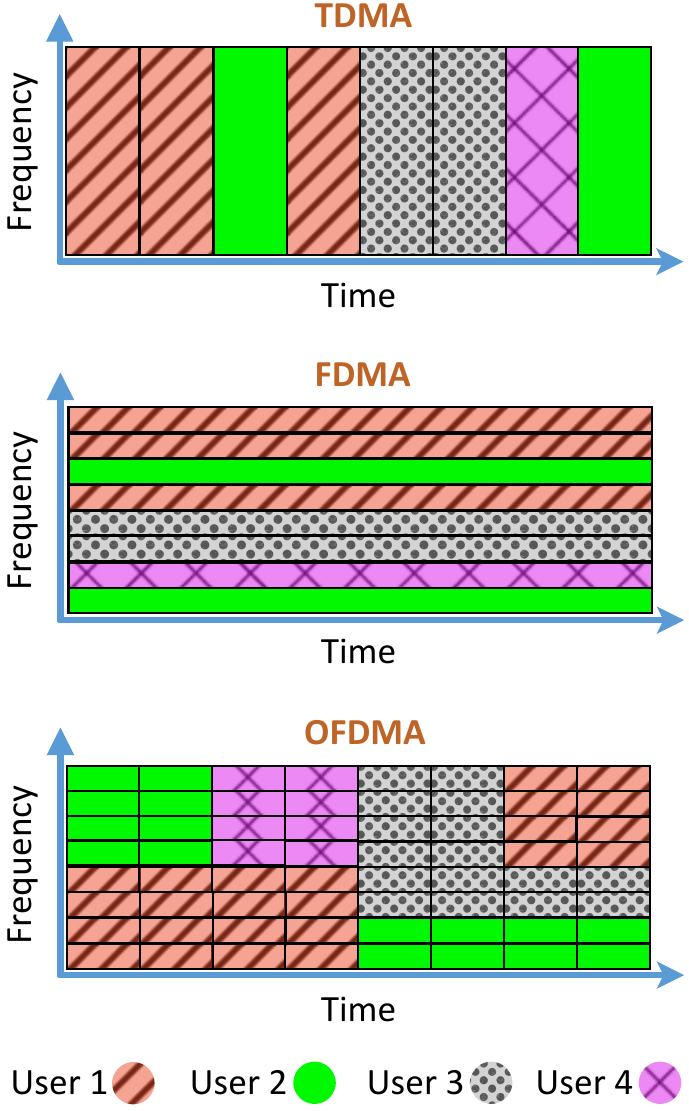}
\end{center}
\caption{Scheduling examples considering that four users are sharing the radio channel.}
\label{fig:multiplexing}
\end{figure}

Distributing the available wireless resources among the users, i.e., multi-user scheduling, is one of the most important tasks that must be implemented in any wireless communication system. Specifically, a scheduler decides how users share the wireless channel by allocating radio resources such as power, time slots, frequency channels, or a combination of these resources.
For instance, Time Division Multiple Access (TDMA) systems are characterized by having time slots as the radio resources units that can be assigned to a user; on the other hand, a scheduler allocates frequency channels in Frequency Division Multiple Access (FDMA) systems; in Orthogonal Frequency-Division Multiple Access (OFDMA) systems,  radio resources are scheduled into the frequency/time domain ---  Fig.~\ref{fig:multiplexing} depicts examples of resource allocation within TDMA, FDMA and OFDMA; for more detail about wireless multiple-access schemes, please refer to \citep{bib:prasad,bib:molisch}.
Nevertheless, from a conceptual point of view, a scheduler can be designed in such a generic way that it is agnostic to which particular radio resources are handled by the underlying wireless multiple-access scheme --- the scheduler only requires knowledge of the total amount of available resource units and the throughput 
provided by each of these units to each of the different users.
As an example, suppose that each resource unit of Fig.~\ref{fig:multiplexing}, within its respective multiple-access scheme, yields the same throughput for all four users, and suppose also that the depicted time/frequency domain span corresponds to a single allocation decision. Accordingly, all three scheduling examples could be generated by the same generic scheduler, namely if the decision was to allocate 3/8, 2/8, 2/8 and 1/8 of the 
maximum achievable system throughput 
to user~1, to user~2, to user~3 and to user~4, respectively. 
 For this reason, some of the proposed wireless resource scheduling techniques follow this generic approach and only point out the percentage of total resources that should be allocated to each user and who should be prioritized, leaving out which specific resources are being handled by the scheduler.

Moreover, designing schedulers for wireless systems comprises many trade-offs among complexity, efficiency and fairness:
\begin{itemize}
\item \emph{Complexity}: It is important to limit the processing time of scheduling algorithms, since they usually have to perform their job under very short periods of time (e.g., 1~ms is the time that Long Term Evolution (LTE) schedulers have for allocation decisions \citep{bib:LTE}). In addition, scheduling schemes should be scalable, meaning that low-complexity algorithms should be preferred over very complex and non-linear solutions, which could be prohibitive in terms of computational cost, time, and memory usage when applied to scenarios with a large number of users.

\item \emph{Efficiency}: Since radio resources are scarce, scheduling algorithms must aim at fully taking advantage of these resources. Performance indicators like the number of users served simultaneously or the average spectral efficiency of the wireless system are two examples of efficiency indicators adopted by many schedulers.


\item \emph{Fairness}: A minimum performance must be also guaranteed for all users, in order to avoid unfair sharing of the wireless resources. Accordingly, implementing the fairness requirement in the scheduling schemes enables that users 
experiencing poor channel conditions (e.g., users that are far away from the base station)
are also served, or that greedy users cannot provoke resource starvation in other users within the same wireless system.

\end{itemize}

In addition to the design factors described above, QoS and QoE provisioning must also be taken into account by the scheduling algorithms. In this section, some scheduling algorithms are reviewed, ranging from the simplest ones to QoS-aware approaches, followed by the introduction of QoS-QoE mapping strategies and utility-based optimization. QoE-based scheduling algorithms are presented in Section~\ref{section:QoE_algs}.

\subsection{QoE-unaware Schedulers}
\label{subsection:non_QoE}

As previously mentioned, any scheduling strategy comprises many trade-offs among complexity, efficiency and fairness. In the case of schedulers that do not take QoE into account, these trade-offs also result from the significance that the different scheduling algorithms give to the communication channel characteristics and to QoS parameters.

\subsubsection{Channel-unaware Strategies} The schedulers that implement these approaches assume that the transmission channel is error-free and time-invariant, which are unrealistic assumptions when dealing with wireless channels. Nevertheless, these strategies form the basis for more complex algorithms.

First In, First Out (FIFO), also known as First Come, First Served (FCFS) \citep{bib:FIFO_RR}, can be regarded as the simplest scheduling scheme, in which users are served according to the order of their resource request. Even though this approach is very easy to implement, it is not fair nor efficient.

The Round-Robin (RR) strategy \citep{bib:FIFO_RR} tries to add some fairness to the FIFO approach, namely by allocating an equal share of resources to each user in a round-robin manner. Thus, this scheduling algorithm is fair regarding the channel occupancy time of each user and can be considered the best choice if the transmitter does not know anything about the channel \citep{bib:molisch}. However, 
RR schedulers are unfair in terms of user throughput because they do not take into account the radio channel conditions (which have a major impact on the throughput).

\subsubsection{Channel-aware / QoS-unaware Strategies} A wireless resource scheduler can take into account the channel state information that is usually fed back to the base stations, so as to enhance the efficiency of its scheduling algorithm.

Maximum throughput (MT) \citep{bib:prasad} is an example of a policy that, in each scheduling period, prioritizes the resources to the user experiencing the best channel conditions. Accordingly, these schedulers provide the highest system throughput, so that the best possible spectral efficiency is attained. Nevertheless, an MT scheduler is very unfair to users with poor channel conditions  and can even make them suffer of starvation.

The concept adopted by Proportional Fair (PF) schedulers \citep{bib:PF_concept} provides a compromise between fairness and spectral efficiency. Within this approach, the average throughput experienced in the past
works as a weighting factor in an MT-like strategy, i.e., if two users can achieve the same throughput (taking into account the channel conditions), then the user that has experienced the lower average throughput is prioritized. This means that users with poor conditions will always be served after some time.

\begin{figure*}[!tb]
\begin{center}
\includegraphics[ width=0.6\textwidth , trim= 0 0 0 0 ]{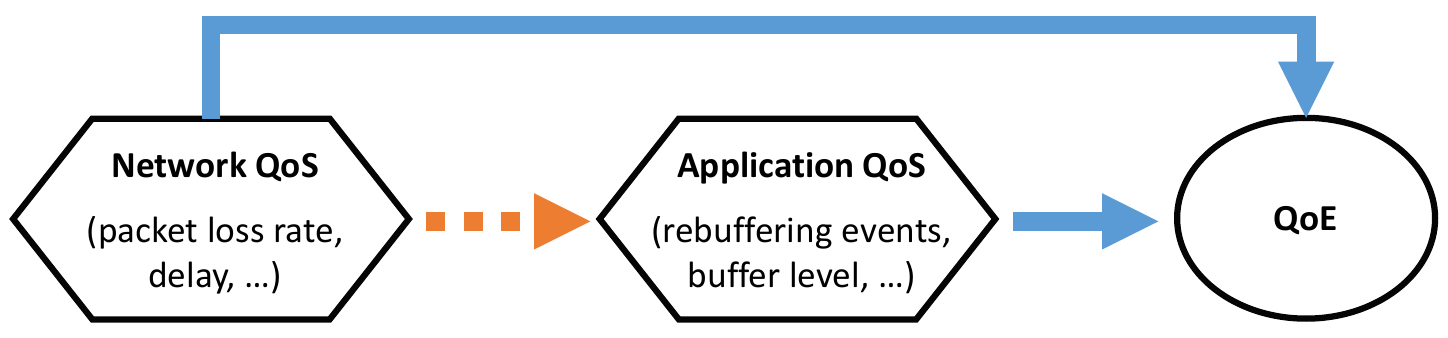}
\end{center}
\caption{Mapping between QoS levels and QoE.}
\label{fig:qos_levels}
\end{figure*}

\subsubsection{Channel-aware / QoS-aware Strategies} For the purpose of attaining a certain performance level, different applications have different requirements, which are typically mapped into QoS parameters. 
Accordingly, scheduling algorithms should also take into account these QoS parameters. For instance, some scheduling strategies try to guarantee a minimum throughput for the users, whereas others deal with delay constrains. This last approach is more common among QoS-aware schedulers, since many applications, such as real-time flows, video streaming or VoIP calls, require that their packets are delivered within a certain deadline.

The Modified Largest Weighted Delay First (M-LWDF) algorithm \citep{bib:M_LWDF} and the Exponential/PF (EXP/PF) scheme \citep{bib:EXP_PF} are two of the most popular QoS-aware scheduling strategies, as they provide a balanced trade-off among fairness, spectral efficiency and QoS provisioning. Besides taking service delay requirements into account, both algorithms support different services and treat differently real-time data flows.

All the above mentioned scheduling strategies, either channel-unaware or channel-aware (with or without taking into account QoS requirements), are just an illustrative sample of what can be found in the literature. The reader is suggested to refer to the surveys by \citet{bib:survey_qos_wimax,bib:survey_qos_1,bib:survey_qos_oppo,bib:survey_qos_lte,bib:survey_qos_2,bib:survey_qos_3} for more examples and details on scheduling algorithms that are not QoE-oriented, namely regarding WiMAX networks, multicast OFDMA systems, opportunistic scheduling, downlink in LTE networks, uplink in LTE and LTE-Advanced, and multi-user Multiple-Input Multiple-Output (MIMO) systems, respectively.

\subsection{QoS-QoE Mapping Strategies}
\label{subsection:QoS_QoE_map}  

QoS-QoE mapping strategies have been presented to quantify QoE, thus making a transition from QoS- to QoE-oriented optimization. QoS-QoE mapping relies on various QoS parameters, which can be divided into two levels: network QoS parameters (e.g., delay or packet loss rate) and application QoS parameters (e.g., rebuffering events or buffer level). Therefore, QoS-QoE mapping strategies try to discover the relationship between QoE and the two QoS levels, where the network QoS parameters are sometimes first mapped into application QoS parameters  --- cf. Fig.~\ref{fig:qos_levels}. Nevertheless, both types of QoS parameters can always be regarded as objective quality metrics, since their measurement is always well defined as they do not depend on any subjective judgment.

Choosing a function $\phi:\mathbb{R}\rightarrow \mathbb{R}$ that establishes a QoS-QoE mapping, i.e., a mapping between the objective quality metrics and a subjective score, is not a straightforward task. 
A linear mapping relationship could be adopted if a certain subjective quality difference always corresponded to the same proportional objective difference \citep{bib:18_of_survey_QoS_QoE_map_1}:
\begin{align}
QoE= \phi_1(QoS_m)=a+b \cdot QoS_m \;, \label{eq:map_linear}
\end{align}
where $a$ and $b$ represent the parameters determined by linear fitting the $m^\text{th}$ objective QoS metric versus the measured subjective scores. 
However, the perceived quality ratings usually do not present a linear behavior with respect to the practical objective quality metrics, meaning that linear mapping functions may lead to an inappropriate assessment of the performance.
To overcome this issue, nonlinear mapping relationships have been adopted, discussed and compared \citep{bib:18_of_survey_QoS_QoE_map_1,bib:survey_QoS_QoE_map_1}; the most widely used can be summarized as follows:
\begin{itemize}
\item Cubic polynomial \citep{bib:20_of_maps_eq_refs,bib:18_of_survey_QoS_QoE_map_1,bib:e_model,bib:wideband_e_model}:
\begin{align}
\phi_2(QoS_m)=a+b \cdot QoS_m + c\cdot QoS_m^2 +d\cdot QoS_m^3 \;.
\end{align}
\item Logistic functions \citep{bib:21_of_maps_eq_refs,bib:18_of_survey_QoS_QoE_map_1,bib:24_of_maps_eq_refs}:
\begin{align}
&\phi_3(QoS_m)=a+\frac{b}{1+c(QoS_m+d)^e} \;, \\
&\phi_4(QoS_m)=a+\frac{b}{1+\exp[c(QoS_m+d)]} \;, \\
&\phi_5(QoS_m)= a + \frac{d}{\left[1+\left(c\cdot{QoS_m}\right)^b\right]^e} \;.
\end{align}
\item Exponential function \citep{bib:23_of_maps_eq_refs,bib:18_of_survey_QoS_QoE_map_1}:
\begin{align}
\phi_6(QoS_m)=a\cdot \exp(b\cdot QoS_m) + c\cdot \exp(d\cdot QoS_m) \;.
\end{align}
\item Power function \citep{bib:18_of_survey_QoS_QoE_map_1}:
\begin{align}
\phi_7(QoS_m)= a\cdot QoS_m^b + c \;.
\end{align}
\item Logarithmic function \citep{bib:22_of_maps_eq_refs}:
\begin{align}
\phi_8(QoS_m)= a\cdot \log(QoS_m + b) +c \;. \label{eq:last_map}
\end{align}
\end{itemize}
Considering the most general case, QoS metrics can be mapped into QoE values by performing a combination of the previous functions \eqref{eq:map_linear}--\eqref{eq:last_map}, including intermediate nonlinear combinations of QoS metrics, i.e.,
\begin{align}
QoE=\sum\limits_i\sum\limits_j\sum\limits_m\phi_i[\phi_j(QoS_m)].
\end{align}

As can be seen, QoE modeling through the use of QoS metrics may encompass complex relationships and interdependencies, with a parametrization that constitutes a non-trivial problem. Moreover, other issues arise when designing QoS-QoE mapping strategies, such as finding out which are the QoS metrics that are more useful for QoE prediction or how much data is needed to achieve a certain accuracy on the estimated QoE. Hence, and in order to tackle these challenges, some authors have proposed the use of machine learning techniques, with the final goal of devising complex models regarding the QoS-QoE relationship \citep{bib:neural_network_qos,bib:view_time,bib:download_ratio_2,bib:download_ratio,bib:qoe_outro_machine,bib:casas}.
 
In all cases, after computing the predicted QoE values, correlation analysis should be carried out between these and ground truth values, so as to assess the goodness of the mapping strategy. On the other hand, traditional QoS optimization techniques can be applied and assessed in QoE optimization scenarios by making use of QoS-QoE mapping strategies. For instance, \citet{bib:20_of_001} investigated and evaluated the performance of three downlink schedulers (PF, M-LWDF and EXP/PF) in terms of QoE metric for VoIP applications over LTE, making it possible to choose the most suitable one in terms of subjective experience.

\subsection{Utility-based Optimization}
\label{subsection:utility_based}  

The concept of utility functions emerged from microeconomics theory and formalizes the relationship between the service performance and the user perceived experience and satisfaction \citep{bib:14_of_TMA_Book}. More specifically, the following utility function $U:X\rightarrow \mathbb{R}$ relates all the resources a user could hypothetically have (set $X$) to real numbers, where $U(x)>U(y)$ indicates that the user has a preference for $x$ over $y$, with $x,\:y \in X$.

The utility-based scheduling optimization may then be regarded as a maximization of the total sum of users' utilities through Network Utility Maximization (NUM) techniques \citep{bib:12_of_TMA_Book}. In mathematical terms, using NUM to allocate network resources (such as transmission power, time slots, etc.) corresponds to perform the following maximization:
\begin{align}
\begin{split}
	&{\text{maximize}} \:\: \sum\limits_{i} U_i\left( x_i\right) \\
	&\text{subject to} \: \sum\limits_{i} x_i \leq X_{\max}
\end{split}, \label{eq:NUM}
\end{align}
where $U_i$ corresponds to the utility function of the $i^\text{th}$ user, $x_i$ stands for the resources allocated to this user, and $X_{\max}$ denotes the bounds of the available resources.

In many cases, scheduling can also be regarded as selecting a throughput vector $\mathbf{R}=\left[ R_1, \ldots, R_K \right]$, for all $K$ users, from the current feasible throughput region $\mathcal{R}$, i.e., the set of achievable throughputs expected for each user according to the respective allocated resources. Thus, the gradient-based scheduling algorithm \citep{bib:gradient} can be applied in order to perform resource allocation decisions:
\begin{align}
		\mathbf{R}^\ast = \argmax_{\mathbf{R}\in \mathcal{R}} \sum_i U'_i(R_i) \cdot R_i , \label{eq:gradient}
\end{align}
where $U'_i$ denotes the derivative of an increasing concave utility function $U_i$ and $R_i$ stands for the achievable throughput expected for the $i^\text{th}$ user. 
For example, the MT scheduler can be obtained from \eqref{eq:gradient} by adopting the utility function $U_i(R_i)={R}_i$, whereas the PF policy derives from the gradient-based scheduling technique with a utility function \mbox{$U_i(R_i)=\log(\overline{R}_i)$}, where $\overline{R}_i$ represents the past average throughput experienced by the $i^\text{th}$ user; accordingly, the selected MT and PF throughput vectors, $\mathbf{R}^\ast_{MT}$ and $\mathbf{R}^\ast_{PF}$ respectively, are given by
\begin{align}
		\mathbf{R}^\ast_{MT} &= \argmax_{\mathbf{R} \in \mathcal{R}} \sum_i {R_i} ,	\\
		\mathbf{R}^\ast_{PF} &= \argmax_{\mathbf{R} \in \mathcal{R}} \sum_i \frac{R_i}{\overline{R}_i} .
\end{align}

\subsection{Discussion}
\label{subsection:disc_background}

Based on what was previously described, QoS-QoE mapping strategies and utility-based optimization can be regarded as the fundamental tools to perform the shift from QoS- to QoE-oriented scheduling. 
Ideally, one should aim at obtaining mathematical formulas that relate application, transport, and physical layer parameters to subjective quality experienced by the users. 
With these, wireless systems design can be adjusted in order to improve the quality perceived by the users. 
For instance, a closed-form expression was presented by \citet{bib:closed_form} concerning the probability of timely transmission of video sequences as a function of the users' allocated bandwidth.
As a consequence, and given a certain QoE requirement based on the probability of timely delivery and the received video stream quality level, the aforementioned expression allows to infer the number of users that can be accommodated in the wireless access system, as well as it can be used to design admission procedures, bandwidth pricing policies, and cell dimensioning. In \citep{bib:qoe_fairness}, a general fairness metric is formulated for shared systems, which satisfies QoE-relevant properties, assuming that estimated QoE values are known. This QoE fairness metric may be adopted when comparing different resource management techniques in terms of their fairness across users and services, although it says nothing about how good the system is and thus needs to be considered together with the achieved (e.g., mean) QoE in system design.

In certain cases, the mathematical formula adopted for a QoS-QoE mapping strategy can also be used as a QoE-oriented utility function, 
namely if there is a well-defined relation between allocation decisions and the considered QoS parameters.
For instance, if a QoS-QoE mapping formula regarding video streaming considers, as single input, the transmitted video bitrate, and assuming that this QoS parameter is directly proportional to the achievable throughput, then a utility function can be derived from this QoS-QoE mapping formula, namely by replacing the QoS input by the corresponding relation between transmitted video bitrate and achievable throughput. 
Another example of a utility function that stems from QoS-QoE mapping strategies can be given regarding file download applications, namely when a QoS-QoE mapping formula only considers, as QoS input, the service response time (i.e., the file download time), which is inversely proportional to the achievable throughput, with a constant of proportionality equal to the size of the file that is being downloaded.

\begin{figure*}[!tb]
\begin{center}
\includegraphics[ width=0.7\textwidth , trim= 0 0 0 0 ]{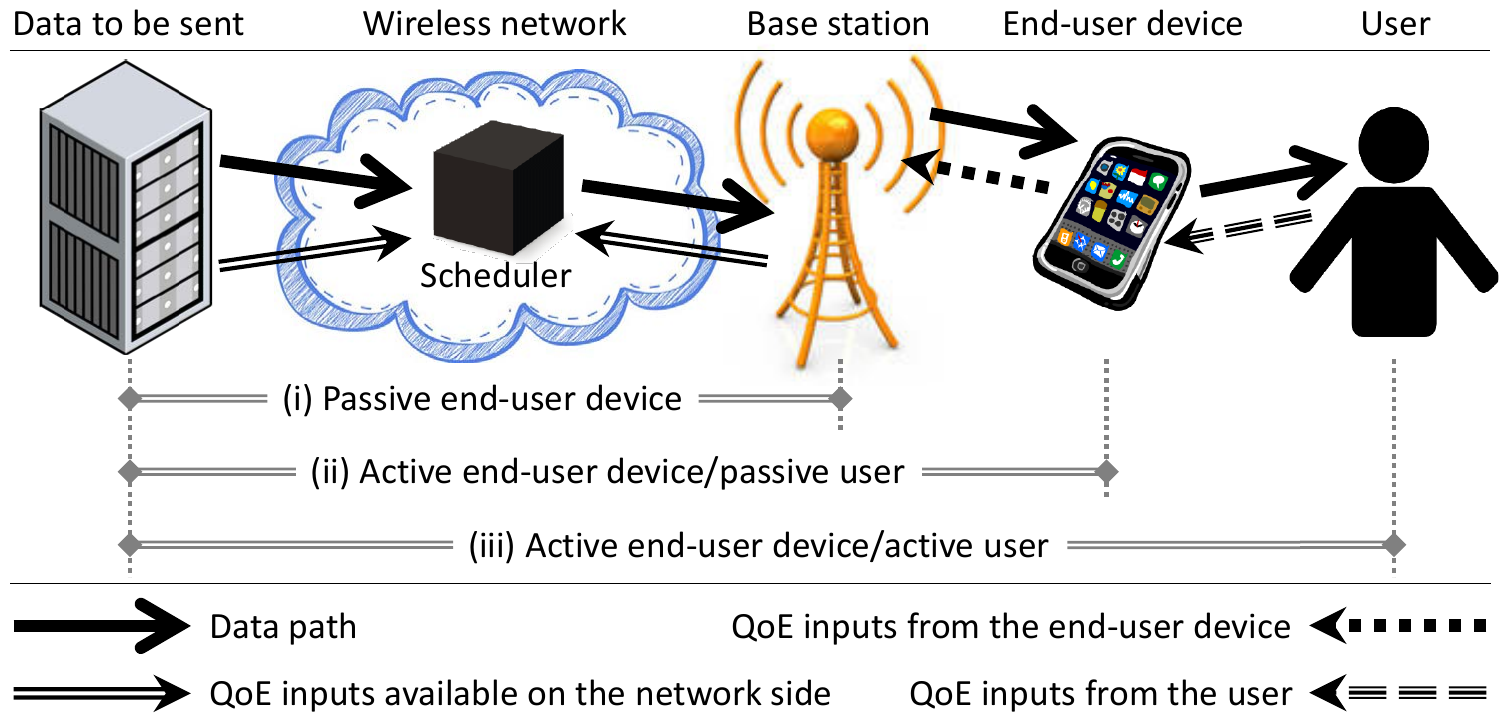}
\end{center}
\caption{QoE-oriented scheduling strategies classification.}
\label{fig:fig_strategies}
\end{figure*}

On the other hand, in many cases, the inputs of QoS-QoE mapping strategies might not have a well-defined relation with the allocation decisions (e.g., when packet loss rate is adopted as QoS input).
Nonetheless, utility functions (or, alternatively, throughput vector selection formulas) can be designed without any knowledge of a specific QoS-QoE mapping strategy and still follow a QoE-oriented approach, as long as the scheduling goals include addressing some issues that affect the users' QoE --- for instance, to try to lessen the impact of packet losses by serving better the respective users afterwards; another example is to perform allocation of resources in order to try to avoid rebuffering events. 
Accordingly, very often QoS-QoE mapping strategies are only required to assess the goodness of a utility function/throughput vector selection formula, i.e., to know the impact of a certain resource allocation policy on the users' QoE. Notice that this last approach can be used to study, in terms of QoE, any scheduling procedure, even those that do not follow a QoE-based design.
For example, and considering video streaming over LTE, the QoE metrics presented in \citep{bib:yaacoub} try to describe the performance of radio resource management methods regarding the end-users subjective video quality. The authors measured minimum, average and geometric mean QoE when scheduling algorithms like MT and PF are adopted.
In \citep{bib:abbas}, the QoE of adaptive video streaming is analyzed under several scheduling policies, like RR and MT, where the QoE is based on the mean video bit rate and the mean buffer surplus. The examination of the performance impact of the different scheduling schemes is then used to suggest the best strategy to be adopted in various mobility scenarios.


\section{QoE-oriented Scheduling Algorithms}
\label{section:QoE_algs}

Many challenges arise when attempting to perform QoE-oriented wireless resources scheduling. As seen in the previous sections, it is important to identify the factors that influence QoE and their relationships to QoE metrics for a given type of service. Some more challenges may be identified after addressing the QoE modeling, namely determining which parameters to collect (e.g., user requirements, network performance, application type, context, etc.), where, how, and when to collect them (e.g., the required parameters could be collected at the base stations or at the end-user devices, either before, during, or after the delivery of the service). Lastly, procedures have also to be defined in order to combine all these steps, i.e., it is necessary to design the methods that allow the collected data to perform QoE-aware scheduling.

In this section, state-of-the-art QoE-based scheduling strategies for wireless systems are reviewed, highlighting the parameters adopted for QoE optimization. To simplify the reading of the survey, the strategies that address the downlink scenario 
have been classified into three categories  --- cf. Fig.~\ref{fig:fig_strategies}: (i) passive end-user device; (ii) active end-user device / passive user; (iii) active end-user device / active user. This classification is based on the adjustments required, at the end-user devices, in order to implement the different scheduling strategies on wireless systems. 
The last part of this section provides a review of QoE-aware scheduling methods that can enhance the wireless resources management in other scenarios, namely the uplink direction, the multi-cell case, under heterogeneous, cognitive radio, relay and multi-user MIMO networks, as well as when dealing with energy-related issues.

\subsection{Passive End-user Device Strategies}
\label{subsection:passive_device}  

Scheduling techniques are easier to implement in a wireless network when the users, as well as their devices, do not perform any exclusive QoE tasks (e.g., monitoring, measuring and reporting relevant parameters), as the QoE assessment is based on measurements that can be carried out solely at the base station side. Since the required assessments can be performed by the scheduler on the network side, no extra information needs to be exchanged between the user's device and the network. On the other hand, these approaches may not achieve the best possible QoE performance, since many relevant metrics, which could be collected at the end-user device (e.g., buffer status), either cannot be used or have to be estimated. 

\subsubsection{Video Streaming}

The simplest QoE-based scheduling approaches consider only the impact of the throughput on the user-perceived quality, namely by adopting the following utility function:
\begin{align}
	U_i(R_i) = f(R_i), \quad f:\mathcal{R} \rightarrow MOS . \label{eq:U_de_rate}
\end{align}
With respect to video streaming, \citet{bib:5_of_15_of_001} and \citet{bib:14_of_001} made use of \eqref{eq:U_de_rate} by first establishing a mapping between video bitrate and MOS, followed by an allocation of resources to each user assuming that the bitrate of the transmitted video is adjusted to match the respective  achievable throughput, so that there is a known correspondence between throughput and MOS.
Besides considering the NUM, it is also proposed in \citep{bib:5_of_15_of_001} another allocation criterion that establishes an a priori target mean MOS of all users, in order to save some network resources (which could be used to serve more users or to support high-demand applications), whereas a tuning mechanism is presented in \citep{bib:14_of_001} that enables the network operator to dynamically adjust the resource allocation between similar perceived quality among all users (system fairness) and maximum average perceived quality (system efficiency). 
\citet{bib:jnca_extra1} proposed a framework to optimize the throughput distribution, which also comprises the determination of video encoding parameters for each user, so that the combined video compression plus radio resource allocation is able to maximize the QoE of all users.
Nevertheless, the previous works neglect the impact of packet loss on the QoE, 
a relevant parameter that was taken into account by other authors. For instance, a trained random neural network is used in \citep{bib:hsdpa} to establish a mapping between the packet loss rate as well as the mean loss burst size and a video QoE score normalized to scale $[0,1]$; next, the respective score of each user, $e_i$, is adopted as a coefficient in modified versions of the MT and PF algorithms, i.e.,
\begin{align}
	\mathbf{R}^\ast = \argmax_{\mathbf{R}\in \mathcal{R}} \sum_i (2-e_i) \cdot w_i \cdot R_i ,
\end{align}
where $w_i$ corresponds to the respective $i^\text{th}$ user weight associated to the MT scheduling policy ($w_i=1$) or the PF one ($w_i=\frac{1}{\overline{R}_i (n)}$). 
In \citep{bib:chinese}, throughput was also considered in conjunction with packet loss rate in the work presented, in which an artificial neural network is adopted to learn the relationship between these parameters and the QoE; afterwards, the scheduler allocates resources based on a particle swarm optimization method, which has the goal of maximizing the users' QoE and, at the same time, balance fairness among them.
\citet{bib:bit_rate} also made use of the utility function \eqref{eq:U_de_rate}, but now throughput is replaced by goodput (i.e., the rate at which the useful data ---  namely excluding retransmitted data packets --- is delivered), so packet loss rate can be considered implicitly, as well as the resource allocation algorithm proposed therein also aims at decreasing the video quality variation. 

The aforementioned scheduling techniques do not take into account the occurrence of playout stalls (which, as mentioned in Section~\ref{subsection:qoe_facts}, are very annoying for the users), mainly because these algorithms assume that the bitrate of the transmitted video is adjusted to match the respective achievable throughput --- hence, in theory, rebuffering events would be avoided. 
However, clients may request video segments with specific bitrates, which means that not only the allocation algorithms must be able to take bitrate constraints as input parameters, but also they should aim at providing interruption-free video transmissions, as playout stalls are more likely to occur if the requested bitrate by a client is too demanding when compared to the respective achievable throughput.
One way to tackle this problem is to consider that the radio resource assignment is given by
\begin{align}
	\mathbf{R}^\ast = \argmax_{\mathbf{R}\in \mathcal{R}} \sum_i \gamma_i \cdot w_i \cdot R_i , \label{eq:U_de_avoid_rebuffering}
\end{align}
where $\gamma_i$ stands for a term that reflects the effect of the $i^\text{th}$ user satisfaction based on the possibility of rebuffering events taking place. 
For instance, \citet{bib:omit_dash} made use of \eqref{eq:U_de_avoid_rebuffering} by defining $\gamma_i$ as an exponential weighted moving average filter that depends on the minimum throughput requirement $R_i^{\min}$ (which stems from the requested video bitrate):
\begin{align}
	\gamma_i  &= e^{\alpha_i^\ast \frac{\alpha_i}{{\|\alpha}\|}} , \\
	\alpha_i  &= \max\left\{ R_i^{\min}  - \overline{R}_i  , \epsilon \right\} ,
\end{align}
where $\alpha_i^\ast$ stands for the $\alpha_i$ value of the previous scheduling period and $\epsilon$ denotes a small positive value. With this approach, a user will be prioritized if the respective achievable throughput does not meet the minimum rate constraint (in order to try to avoid playout stalls), whereas if this target is met, then the user experiences a very low $\gamma_i$ weight, thus being deferred from being served.
Another approach based on  \eqref{eq:U_de_avoid_rebuffering} is proposed in \citep{bib:flexible}, namely by considering $w_i=1$ and a $\gamma_i$ weight given by
\begin{align}
	\gamma_i = \left( 1-\frac{R_i}{R_i^{\min}} \right)^\beta ,
\end{align}
where $\beta$ corresponds to a value that can be adjusted to achieve a certain trade-off between fairness (high $\beta$ values) and efficiency ($\beta$ close to zero).

Still regarding the occurrence of playout stalls, \citet{bib:lower_bound} proposed a lower bound for a mean rebuffering percentage (percentage of the entire streaming time in which a user is experiencing playout stalls), along with a corresponding optimal scheduling strategy. Their approach yields, for each user, an achievable throughput value that ranges from zero to $R_i=R_i^{\min}$ --- more precisely, the scheduler serves the users that require less resources (in order to fulfill their $R_i^{\min}$ demand) until there are no more resources available, thus meaning that users with a high $\frac{R_i^{\min}}{C_i}$ relation are more prone to be left out, where $C_i$ represents the maximum achievable throughput if all resources were assigned to user $i$.
Nevertheless, this scheduling technique is not suitable for the case where all users can be served at the same time without network congestion, namely because some spare resources would not be allocated (which could lead to an increase of the initial playout delay). In addition, the aforementioned lower bound assumes that the video representation quality chosen at the beginning will be the same throughout the whole streaming session, which is not true if adaptive video streaming is enforced.

Another factor that may influence the users' QoE, and which has not been addressed by the previous scheduling algorithms, is when one or more subscribers should be prioritized over the remaining because, e.g., they are paying more in order to obtain a better service. One way to tackle this issue is to divide the users into different classes and assign different priority weights to them. This approach was followed by \citet{bib:heavy}, in conjunction with a scheduling technique that tries to minimize the duration of playout stalls. 
More specifically, it is proposed to schedule the client with the largest $R_i$ in each scheduling period and, if a tie occurs, the chosen client is the one that has the smallest $\nu_i \left( \overline{R}_i - R_i^{\min} \right)$, where $\nu_i$ corresponds to a predetermined weight that takes into account the user's  class. In order to compute $\nu_i$, the authors adopt the NUM technique and obtain some tractable solutions; however, it is important to mention that the scheduling technique presented in \citep{bib:heavy} is designed assuming some conditions, namely that the sum of the minimum throughputs required by the users is not higher than the maximum achievable system throughput. Moreover, since a user will always be sacrificed if the respective channel conditions are poorer than the ones of another user, this scheduling policy is more suitable in scenarios where the throughput of the wireless link is expected to be similar among all users.

Some authors have also considered another relevant parameter not addressed so far, namely the Head-of-Line (HoL) packet delay (i.e., the delay of the first packet), in order to perform QoE-oriented scheduling that is somewhat capable of minimizing playout stalls. In \citep{bib:video_delay}, a modified version of the PF algorithm is proposed, which adopts the following scheduling decision:
\begin{align}
	\mathbf{R}^\ast = \argmax_{\mathbf{R}\in \mathcal{R}} \sum_i  \frac{R_i}{\overline{R}_i} +  a \cdot \exp\{b(\tau_i-\overline{\tau}_i)\}\frac{R_i^{\min}}{R_i}  , 
\end{align}
where $\tau_i$ and $\overline{\tau}_i$ denote the HoL packet delay and the average packet delay, respectively, regarding the $i^\text{th}$ user, whereas $a$ and $b$ represent some constants which enable to adjust the impact of the delay variables on the scheduling decision.
%
An approach based on the M-LWDF technique is proposed in \citep{bib:802_11}, in which the scheduling process jointly considers packet delay, expected throughput and video importance. More specifically, before scheduling a user, some overdue packets are discarded within this scheme, namely those with an associated delay that has exceeded the deadline threshold given by $\sqrt{^{p_i^{(2)}}/_{p_i^{(1)}}}$, where $p_i^{(1)}$ and $p_i^{(2)}$ correspond to positive tuning parameters that enable to adjust the deadline threshold with respect to the $i^\text{th}$ user. Afterwards, radio resource assignment is performed by following the scheduling rule given by
\begin{align}
	\mathbf{R}^\ast = \argmax_{\mathbf{R}\in \mathcal{R}} \sum_i  p_i^{(1)}  \cdot I_i \cdot \tau_i \cdot R_i  , 
\end{align}
where $I_i$ stands for the video importance index of the packet that is being download by the $i^\text{th}$ user, which is derived from an algorithm described in \citep{bib:802_11} and aims at setting a value for how important is a packet in enhancing the video quality. 
\citet{bib:recent} proposed a scheduling policy that is similar to the previous one, in the sense that not only it takes into account packet delay, expected throughput and video importance, but also the scheduler is allowed to discard some packets. However, the packet filtering process is now based on the respective contribution towards video quality (instead of overdue packets), under an algorithm that is adjusted to work with scalable video streaming. The scheduling decision of this method is as follows:
\begin{align}
	\mathbf{R}^\ast = \argmax_{\mathbf{R}\in \mathcal{R}} \sum_i  \exp\{ [P_i]^{\overline{\tau}} \cdot \tau_i  \}\frac{R_i}{\overline{R}_i}  q_i  , 
\end{align}
where $P_i$ corresponds to the priority rating of the packet that is being downloaded by the $i^\text{th}$ user, which takes into account the respective bitrate and contribution towards the perceived video quality, and $q_i$ denotes the number of packets currently residing in the queue to be scheduled to user $i$.
It is noteworthy to stress that these last two scheduling techniques are not lossless, i.e., a client may not receive all the information it asked for; on the other hand, resources can be saved (and further allocated in order to enhance the QoE) by not transmitting overdue packets or with low contribution to the video quality, which might have become useless at the receiver in the sense that they would not provide a great QoE enhancement.

\subsubsection{Other Applications}

The user experience in some wireless applications other than video streaming can also be enhanced by the use of QoE-aware scheduling methods.
With respect to VoIP, the algorithm presented in \citep{bib:voip_delay} tries to allocate resources in order to limit the delay within a certain deadline and, consequently, meet the tight delay requirements of VoIP; in addition, the proposed framework also has the goal of minimizing the total number of radio resources scheduled during a certain period of time, a procedure which is based on not serving some users at some time instances if their forecasted channel conditions are able to cope with future transmissions that still meet the deadline.  
\citet{bib:7_of_15_of_001} addressed web browsing applications and suggested a mapping from user throughput to user experienced quality, which enables to perform radio resource allocation through the maximization of the aggregate utility over all users; the respective utility function is given by
\begin{align}
	U_i(R_i)=5-\frac{578}{1+\left( 11.77+ 22.61\cdot \frac{R_i}{W_{S}} \right)^2} ,
\end{align}
where $W_{S}$ stands for the Web page size --- note that this approach can also be regarded as one that aims at rendering the Web page with a service latency lower than approximately 10 seconds, namely because the ratio $\frac{W_{S}}{R_i}$ is intended to correspond to the service response time measured in seconds.

In the general case, wireless networks provide multi-services, where video streaming, VoIP, Web browsing and file download applications are available to the users; consequently, there should be a scheduling concern of providing high QoE for all of them. For instance, the scheduler proposed in \citep{bib:beauty} tries to maintain the past average throughput values per user, in order to moderate throughput fluctuations; more specifically, the authors adopt the following scheduling rule:
\begin{align}
	\mathbf{R}^\ast = \argmax_{\mathbf{R}\in \mathcal{R}} \sum_i  \frac{1}{\overline{R}_i - R_i}  .
\end{align}
Noticing that this approach is application-unaware, it has the advantage that there is no need for the scheduler to know which service each user is using. On the other hand, and since QoE is also application-dependent (as mentioned in Section~\ref{subsection:qoe_facts}), 
scheduling strategies should be adjusted in order to take into account the particularities of each service. For instance, the work presented in \citep{bib:multi_tudo} proposes a NUM-based scheduling scheme, in which different utilities functions are adopted for each service --- the authors consider mapping functions based on throughput plus packet loss regarding video streaming, the delay is regarded as the relevant parameter for VoIP, whereas the QoE of Web browsing and file download applications are based on service response time and throughput, respectively. This scheduling algorithm was assessed for two operation modes, namely one that aims at maximizing the sum of all users' QoE and another with the optimization target of maximizing the sum of the logarithm of the users' QoE --- it is claimed that the latter mode improves the fairness among services without a great impact on the average QoE.
The resource allocation scheme described in \citep{bib:sug_2} also aims at maximizing the average QoE regarding multi-services; the authors devised a personalized strategy, in which QoE is evaluated not only using QoS factors, such as throughput, packet loss rate or delay, but also considering a predicted user preference based on contextual factors (e.g., the registered age, gender and occupation of the user, time of the day, day of the week, duration of the content and its popularity, etc.). 
%
\citet{bib:anand_veciana} designed a scheduling method which takes into consideration the mean flow delays in multi-service systems. More specifically, they propose a framework based on the Gittins index to solve the optimization problem given by
\begin{align}
	\inf_{\mathbf{d}^\pi}  \left\{ \sum_s \lambda_s f_s(d_s) \: | \: \mathbf{d}^\pi \in \mathcal{D}  \right\} ,
\end{align} 
where the $\mathbf{d}^\pi=[d_1^\pi,\ldots ,d_S^\pi]$ denotes the mean delay vector realized by the scheduling policy $\pi$ (for all $S$ services) from the feasible delay region $\mathcal{D}$, i.e., the set of possible mean delay vectors considering all policies; with respect to service $s$, $\lambda_s$ stands for the arrival rate, $d_s$ represents the mean delay experienced and $f_s(\cdot)$ corresponds to the cost function that reflects the respective QoE sensitivity.

In \citep{bib:sacchi_draft}, a wireless resources assignment method is presented that also has the goal of providing a similar QoE among all users (which have miscellaneous requirements for video, voice and data services), namely by making use of game theory concepts, as well as minimum throughput requirement and packet loss probability of the different services, in order to maximize the minimum QoE.
A scheduling procedure is proposed in \citep{bib:messaging} with respect to instant messaging service, which can incorporate video chat, audio chat and text chat subservices, and where a user may launch several subservices at the same time (e.g., multiple text chats and an audio chat). More precisely, the authors adopt the average delay of image, voice and text flows as the base metric to quantify QoE, along with a method in which the scheduler computes, without feedback from the terminals, the probability that a user is focusing on a certain subservice (taking into account the subservice type and its serving quality), thus regarding the one with the higher probability as the representative service. Based on this premise, the scheduling scheme first looks up for the user experiencing the poorest QoE of the representative service and then allocates resources for one subservice of this user; regarding this last step, a random approach is devised, such that even though the representative service has a higher chance of being prioritized, the other subservices will not be starved (which also helps to handle inaccurate estimations of the representative service). 
%


It is important to stress that even though the previous approaches aim at ensuring that all users have an identical QoE (system fairness), this might lead to undesirable situations: for instance, when one user is requesting a very demanding service or is experiencing very poor channel conditions, a QoE-fair scheduler would allocate more resources to this user and eventually force the majority of the remaining users to have a poor experience. On the other hand, a scheduler that maximizes the average perceived quality (system efficiency) could also sacrifice some users by not providing them, at least, an acceptable QoE. For this reason, and regarding multi-service systems, some works proposed solutions that try to offer a trade-off between system fairness and system efficiency \citep{bib:gao,bib:xing_fei_li,bib:monteiro,bib:RRA,bib:hori} --- although these works are somewhat similar, their differences deserve to be highlighted, which will be done in the following paragraph.

According to the scheme introduced in \citep{bib:gao}, the users are served following an MT fashion until all of them have the respective minimum throughput requirement satisfied (it is assumed that QoE is estimated by taking into account solely the throughput of video, audio and file download applications); in the following step, the resources are allocated to the users that can achieve the best QoE gain. 
In \citep{bib:xing_fei_li}, the proposed scheduler performs the same first step as the previous solution, i.e., users are served following an MT fashion until all of them have the respective minimum throughput requirement satisfied; afterwards, the remaining resources are allocated in a fair manner such that almost the same quantity is assigned to all users.
In \citep{bib:monteiro}, the proposed algorithm serves the users following an MT fashion until a predefined number of users is satisfied by having a QoE equal or greater than a certain threshold (it is assumed that throughput can be mapped into QoE); afterwards, the remaining resources are allocated to the users with the lowest QoE.
\citet{bib:RRA} presented a method that searches for the users with the minimum QoE and assigns resources to them; this process is repeated until each user is considered satisfied according to the respective minimum QoE (the authors adopted mapping functions based on throughput plus packet loss regarding video streaming and audio applications, whereas for Web browsing applications the considered relevant parameter was service response time); next, the remaining resources are allocated following an MT fashion, with the proviso that users that achieve a very high satisfaction threshold are excluded from being further served. 
The approach of \citet{bib:hori} is very similar to the previous one, with the main difference that all users are considered satisfied according to the same QoE threshold (the adopted mapping functions are also slightly different, namely delay is now used as relevant parameters for audio applications); this criterion might be less realistic than the previous one, in the sense that users often expect a different level of satisfaction for different services.

\begin{table*} [!tb] 
\footnotesize
\renewcommand{\arraystretch}{1.3}
\centering
\caption{Passive End-user Device Strategies.}
\vspace{8pt}

\begin{tabular}{| c || c || c  | c | c || c |} 
\hline
\multirow{2}{*}{\textbf{Application}} & \multirow{2}{*}{\textbf{Reference}} & \multicolumn{3}{|c||}{\textbf{QoE optimization based on}} & \multirow{2}{*}{\textbf{Additional comments}} \\  
\cline{3-5}	& & Throughput & Packet loss rate & Delay  & \\
\hline\hline

\multirow{12}{*}{\makecell{Video\\streaming}} 	&  \citet{bib:5_of_15_of_001} & X &  & & \multirow{6}{4.1cm}{{Video bitrate is assumed to\\ be adjusted to match the achievable throughput. }} \\ 
\cline{2-5}	
	&  \citet{bib:14_of_001} & X &  & & \\ 
\cline{2-5}	
	&  \citet{bib:jnca_extra1} & X &  & & \\ 
\cline{2-5}	
	&  \citet{bib:hsdpa} & X & X & &  \\ 
\cline{2-5}	
	&  \citet{bib:chinese} & X & X & &  \\ 
\cline{2-5}	
	&  \citet{bib:bit_rate} & X & X & &  \\ 
\cline{2-6}	
	&  \citet{bib:omit_dash} & X &  & & \multirow{7}{4.1cm}{{Video bitrate constraints are taken as  input parameters.}} \\ 
\cline{2-5}	
	&  \citet{bib:flexible} & X &  & &  \\ 
\cline{2-5}	
	&  \citet{bib:lower_bound} & X &  & &  \\ 
\cline{2-5}	
	&  \citet{bib:heavy} & X &  & &  \\ 
\cline{2-5}	
	&  \citet{bib:video_delay} & X &  & X &  \\ 
\cline{2-5}	
	&  \citet{bib:802_11} & X &  & X &  \\ 
\cline{2-5}	
	&  \citet{bib:recent} & X &  & X &  \\ 
\hline

\multirow{1}{*}{{VoIP}} 	&  \citet{bib:voip_delay} &  &  & X & \multirow{1}{4.1cm}{{  }} \\ 
\hline

\multirow{1}{*}{{Web browsing}} 	&  \citet{bib:7_of_15_of_001} &  &  & X & \multirow{1}{4.1cm}{{  }} \\ 
\hline

\multirow{14}{*}{{Multi-service}} 	&  \citet{bib:beauty} & X &  &  & \multirow{1}{4.1cm}{{Application-unaware solution. }} \\ 
\cline{2-6}	
	&  \citet{bib:multi_tudo} & X & X & X & \multirow{3}{4.1cm}{{Aim at maximizing the average QoE. }} \\  
\cline{2-5}
	&  \citet{bib:sug_2} & X & X & X & \\
\cline{2-5}
	&  \citet{bib:anand_veciana} &  &  & X & \\	
\cline{2-6}	
	&  \citet{bib:sacchi_draft} & X & X & &  \multirow{2}{4.1cm}{{Attempt to provide similar QoE. }}\\  
\cline{2-5}	
	&  \citet{bib:messaging} &  &  & X & \\ 
\cline{2-6}	
	&  \citet{bib:gao} & X &  & &  \multirow{5}{4.1cm}{{Try to offer a trade-off between providing similar QoE and maximizing the average QoE. }} \\ 
\cline{2-5}	
	&  \citet{bib:xing_fei_li} & X &  & &  \\ 
\cline{2-5}	
	&  \citet{bib:monteiro} & X &  &  & \\ 
\cline{2-5}	
	&  \citet{bib:RRA} & X & X & X &  \\ 
\cline{2-5}	
	&  \citet{bib:hori} & X & X & X &  \\ 
\cline{2-6}	
	&  \citet{bib:natal_2} & X &  &  &  \multirow{2}{4.1cm}{Prioritize video streaming.} \\ 
\cline{2-5}	
	&  \citet{bib:extra_2019_c} &  \multicolumn{3}{c||}{Buffer estimation through packet inspection.} &  \\ 
\hline

\end{tabular}
\label{tab:passive_device}
\end{table*}

More recently, \citet{bib:natal_2} proposed a scheduling strategy that prioritizes video streaming over other applications; in particular, resources are allocated to video users according to the policy
\begin{align}
	\mathbf{R}^\ast = \argmax_{\mathbf{R}\in \mathcal{R}} \sum_i  (R_i^{\min} - {\overline{R}_i})\cdot R_i ;
\end{align}
however, if the average throughput of all the video flows is above their respective throughput requirement (i.e., $\forall_i:{\overline{R}_i} > R_i^{\min}$), then the spare resources are assigned to non-video traffic, thus improving the system utilization.
The work of \citet{bib:extra_2019_c} also treats video delivery in a special way, namely by using real-time network-based machine learning classifiers, which make use of standard unencrypted packet headers in order to not only detect the service type of different flows, but also to estimate the player status regarding video streaming users; the authors claim that their framework is able to provide an enhanced video QoE (with an acceptable impact on other non-video services), namely by increasing the prioritization weight of users detected as being experiencing a rebuffering event.

\subsubsection{General Considerations on Passive End-user Device Strategies}

The previous overview demonstrates that the research community has made some efforts to present scheduling methodologies that enhance the QoE of the users and which rely on measurements that can be carried out solely at the base station side --- these strategies are summarized in Table~\ref{tab:passive_device}. In accordance with the popularity achieved by video streaming services over the last years, the majority of the works proposed QoE-oriented scheduling solutions taking into account this type of service. Moreover, many of these works aim at allocating resources with the goal of proving interruption-free video; however, their approach may still be inefficient: for instance, consider a user that, at a certain scheduling period, has a large amount of data stored in the respective buffer --- e.g., the user had paused the playback for a while and, in the meantime, the buffer stored some dozens of seconds of video; 
if, by following the previously mentioned algorithms, this user is prioritized, then this scheduling decision occurs in a situation where the respective video could be played smoothly during some time, even if this user was not served. Accordingly, this user is being favored over others that might have nearly-empty buffers and, if not served, will experience playout stalls.

Many proposals also addressed the multi-service case, in which the adopted objective functions have a dual role: not only they have to capture the particularities of each service regarding QoE by using relevant metrics that enable to accomplish that goal (e.g., throughput plays a central role when designing QoE-oriented schedulers for the video streaming service, whereas delay is of prime importance for VoIP applications), but also the different objective functions establish a trade-off regarding the different services, namely which service should be prioritized and when this should occur. For instance, users that are performing web browsing or large file downloads may not perceive a QoE degradation if the respective service delay is slightly increased; hence, considering a congested system, it may be beneficial to prioritize applications such as video streaming or VoIP, so that their QoE is not affected or is even increased. On the other hand, considering uncongested systems, scheduling more resources to the latter applications would enhance their QoE only marginally if these applications are already properly served; in this situation, it would make more sense to allocate the remaining resources to web browsing or large file download applications. 
Therefore, the choice of objective functions that are able to meet simultaneously the two aforementioned goals turns multi-service QoE-oriented scheduling in a tougher challenge when compared to single-service systems. In addition, there is also another issue that usually arises within multi-service systems, namely the need for the scheduler to know which service each user is using, a requirement which either involves detecting the service type through packet inspection 
or entails an extra cooperation between the application-level server and the scheduler.


\subsection{Active End-user Device / Passive User Strategies}
\label{subsection:passive_user}  

One way to enhance the subjective quality perception of a service is to perform QoE assessment as close as possible to the end-user and to report this information to the base station. In this way, there is a higher degree of confidence regarding the influence of the scheduler adjustments versus the QoE improvement, since more accurate QoE metrics can be used. However, this type of scheduling algorithms require feedback channels for QoE measurements reporting.
For instance, the Dynamic Adaptive Streaming over HyperText Transfer Protocol (DASH) technology, also known as MPEG-DASH \citep{bib:dash_mpeg}, and its extensions standardized by the 3\textsuperscript{rd} Generation Partnership Project (3GPP) for DASH use over wireless networks (3GP-DASH)  \citep{bib:dash_3gpp} already incorporate specifications about how relevant QoE parameters (e.g., video buffer level) can be reported to the server --- it is noteworthy to mention that one of the goals of this technology is to cope with variable network conditions, such as those caused by wireless link quality variations, namely by defining how different representations (with different bitrates) of the same multimedia content can be split into smaller segments, which can be independently decoded by a client, and how the clients can select and retrieve these segments from a DASH server; accordingly, a DASH client is able to switch seamlessly between different representations during a streaming session, an adaptation which, for example, enables to minimize the impact of throughput fluctuations on playout stalls, thus enhancing the QoE.


In the scheduling strategies described in this subsection, the end-users devices are regarded as clients that are able to measure and to report to the base stations the QoE metrics that will be be taken into account by the scheduler, while the user itself does not perform any action.

\subsubsection{Video Streaming}

\begin{figure}
\centering
\subfloat[Signaling events.]{\includegraphics[width=0.4\columnwidth,trim={0 0 0 0},clip]{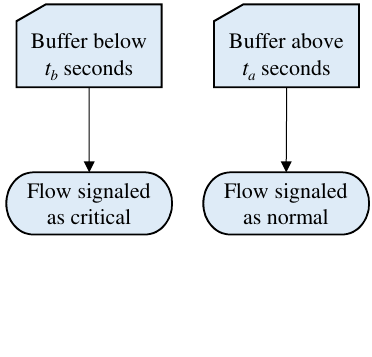}%
\label{fig:signal_events}}
\hfil
\subfloat[Scheduling procedure.]{\includegraphics[width=0.6\columnwidth,trim={0 0 0 0},clip]{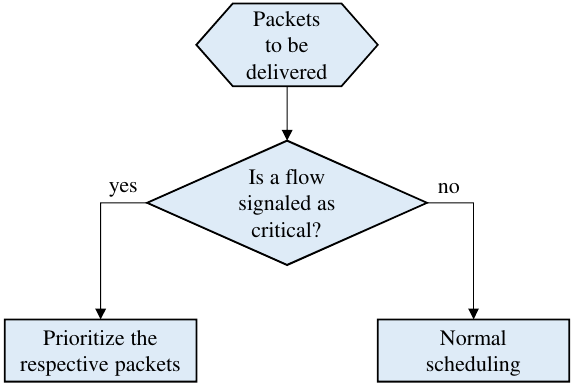}%
\label{fig:sched_proc}}
\caption{Flowchart of an active end-user device / passive user scheduling algorithm for video streaming that runs on top of traditional schedulers \citep{bib:16_of_001,bib:no_copy}.}
\label{fig:sched_flow}
\end{figure}

The buffered playout time (at the end-users devices) of YouTube videos is considered in \citep{bib:16_of_001,bib:no_copy} in order to generate signaling events when the buffered video playout time drops below threshold $t_b$ or goes above threshold $t_a$ ($t_a \geq t_b$), cf. Fig.~\ref{fig:sched_flow}: in the former case, video flows are tagged as being in a critical state, thus the scheduler prioritizes the respective packets; when the buffered playout time of a critical video flow becomes greater than $t_a$, it is relabeled as a normal flow and the scheduler allocates resources according to any scheduling policy (e.g., MT, PF, etc.). This methodology, which is intended to run on top of traditional scheduling algorithms, is not proactive regarding overall QoE improvement, i.e., instead of aiming at QoE amelioration for all users at all scheduling instances, this scheme only prioritizes video flows when QoE impairment, namely a playout stall, might be imminent as a result of a buffer that is becoming empty.
Another scheduling method that attempts to avoid QoE degradation is presented in \citep{bib:playout_lead}, in which the goal is to maximize the minimum buffered video playout time among all users.
On the other hand, a more proactive approach can be adopted, in order to also try to enhance the users' overall QoE. For instance, \citet{bib:001} proposed a scheduler that considers $QoE \propto \frac{1}{D_{MOS}}$, where $D_{MOS}$ is a positive value that denotes the overall perceived quality deterioration, which is not only based on ongoing rebuffering events, but also takes into account past playout stalls; accordingly, and in order to provide a fair QoE, the user with the highest $D_{MOS}$ is prioritized in each scheduling period --- still regarding the fairness issue, a protection mechanism is also conceived so that a user is not continuously prioritized over the others for more than a certain amount of time.

In general, the aforementioned scheduling techniques do not take advantage of the higher throughputs that can be attained by the users that are experiencing better channel conditions --- if these users are better served, then the average perceived quality (system efficiency) may be improved, provided that the users that are experiencing poor channel conditions can still achieve a satisfactory QoE. Hence, some scheduling procedures have been devised that make use of the reported buffer level and aim at increasing the system efficiency regarding QoE.
In \citep{bib:NOVA}, the following scheduling rule is proposed for radio resource assignment:
\begin{align}
	\mathbf{R}^\ast = \argmax_{\mathbf{R}\in \mathcal{R}} \sum_i h_i^v(v_i)  \cdot R_i  , 
\end{align}
where $v_i$ stands for an indicator of risk of violation of rebuffering constrains by the $i^\text{th}$ user and $h_i^v(\cdot)$ denotes a non-negative Lipschitz continuous function such that $\lim_{v \rightarrow \infty} h_i^v(v)=\infty$, $h_i^v(v_i)=0$ for all $v_i \leq \underline{v}$ for some constant $\underline{v}$ (typically set as zero), and is strictly increasing for $v_i \geq \underline{v}$ --- the authors suggest an algorithm in which the parameter $v_i$ is updated, in each scheduling period, symmetrically with respect to the buffered playout time of user $i$, denoted from now on as $B_i$, i.e., $v_i$ is roughly a linear decreasing function of  $B_i$.
Some proposals have also been presented which are based on the PF technique, 
in the sense that they follow the PF scheduling behavior but, whenever the buffer level of some users starts running low, they also carry out a smooth behavior transition in order to allocate more resources to these users.
For instance, the scheduling technique introduced in \citep{bib:15_of_001} --- which was also adopted for  multicast systems by \citet{bib:multicast} --- aims at prioritizing users with low buffer levels, with the proviso that these users will not consume an excessive amount of resources; this goal is pursued by the scheduling rule
\begin{align}
	& \mathbf{R}^\ast = \argmax_{\mathbf{R}\in \mathcal{R}} \sum_i \frac{R_i}{\overline{R}_i} L(B_i),
\end{align}
where $L(\cdot)$ denotes a decreasing logistic function --- hence $L(B_i)$ is upper bounded for low $B_i$, thus avoiding a monopolization of the resources by one or more users with low buffer levels.
In \citep{bib:oyman}, the variation rate of the buffer level is also taken into account when performing resource allocation, with the purpose of a continuous adjustment of the scheduling priorities in order to prevent low buffer levels in the first place; the corresponding resource allocation decision is as follows:
\begin{align}
 \mathbf{R}^\ast = \argmax_{\mathbf{R}\in \mathcal{R}} \sum_i \frac{R_i}{\overline{R}_i} \exp\left\{ \eta_i \cdot T_i \right\},
\end{align}
where $\eta_i$ represents a parameter that determines the time-scale over which rebuffering constrains are enforced for the $i^\text{th}$ user --- this parameter reflects the current buffer level, in the sense that the authors suggest an algorithm that scales $\eta_i$ to prioritize users with low buffer levels --- and $T_i$ corresponds to a video-aware user token parameter, which should increase or decrease whenever the variation rate of the buffer level is below or above a certain threshold, respectively.
The strategy proposed in \citep{bib:buffered_video} also tries to provide fairness in terms of rebuffering percentage (i.e., the percentage of the total streaming time spent rebuffering); accordingly, the respective scheduling decision is given by
\begin{align}
	& \mathbf{R}^\ast = \argmax_{\mathbf{R}\in \mathcal{R}} \sum_i \left(\frac{R_i}{\overline{R}_i} +  \frac{\mu\cdot R_i}{F_i} \exp\left\{ \xi (f_{\min} - f_i)\right\} \right) V_i , \\
  & V_i = \begin{cases} 
1+\frac{N_U\times p_{rebuf,i}}{\sum_{j=1}^{N_U} p_{rebuf,i}}, & \text{if } \sum_{j=1}^{N_U} p_{rebuf,i}>0 \\
1,   & \text{otherwise}
\end{cases} ,
\end{align}
where $F_i$ corresponds to the size of the video frame that the $i^\text{th}$ user is downloading, $f_i$ stands for the number of video frames in the buffer of this user, and $f_{\min}$, which is tunable, represents the minimum number of frames that each client should have in the buffer;  
the variable $p_{rebuf_i}$ denotes the rebuffering percentage of the $i^\text{th}$ user, $N_U$ corresponds to the total number of connected users, whereas $\mu$ and $\xi$ stand for other tunable parameters of this scheduling algorithm. 

It is noteworthy to mention that even though the previous scheduling strategies exploit the higher throughputs that can be attained by some users, this approach might still be inefficient --- for instance, prioritizing the user that is experiencing the highest throughput may not imply that the respective buffer level will have a great increment (e.g., this user could be requesting a very high quality video); therefore, it might be more efficient to serve other users which can store a higher amount of playout time in their buffers, in order to better obviate the occurrence of stalls. This approach is followed by the scheduling algorithm introduced in \citep{bib:fred}, namely by considering explicitly the predicted buffer level variation of the user $i$, $\Delta B_i$, according to the resources that could be allocated to this user:
\begin{align}
\mathbf{R}^\ast = \begin{cases} 
\displaystyle\argmax_{\mathbf{R}\in \mathcal{R}} \sum_i \frac{1}{B_i}, & \text{if } \exists j: B_j < \Omega \\
\displaystyle\argmax_{\mathbf{R}\in \mathcal{R}} \sum_i \frac{\Delta B_i}{B_i},   & \text{otherwise}
\end{cases} ,
\end{align}
where $\Omega$ denotes an emergency state threshold (somewhat equivalent to the critical state of Fig.~\ref{fig:sched_flow}), which is activated when one or more users have a buffered playout time lower than $\Omega$, i.e., this emergency condition is useful in preventing some users from suffering from starvation (a situation which would give rise to playout stalls).

\begin{table*} [!tb] 
\footnotesize
\renewcommand{\arraystretch}{1.3}
\centering
\caption{Active End-user Device / Passive User Strategies.}
\vspace{8pt}

\begin{tabular}{| c || c ||  c | c || c |} 
\hline
\multirow{2}{*}{\textbf{Application}} & \multirow{2}{*}{\textbf{Reference}} & \multicolumn{2}{|c||}{\textbf{QoE optimization based on}} & \multirow{2}{*}{\textbf{Additional comments}} \\  
\cline{3-4}	& & Buffer level & Other  & \\
\hline\hline

\multirow{14}{*}{{\makecell{Video\\streaming}}} 	&  \citet{bib:16_of_001} & X &  & \multirow{2}{4.1cm}{{Run on top of existing schedulers (less  proactive). }} \\ 
\cline{2-4}	
	&   \citet{bib:no_copy} & X &  & \\ 
\cline{2-5}	
	&   \citet{bib:playout_lead} & X &  & \multirow{2}{4.1cm}{{Attempt to offer similar QoE. }} \\  
\cline{2-4}	
	&   \citet{bib:001} &   & Playout stalls &  \\ 
\cline{2-5}	
	&   \citet{bib:15_of_001} &  X & & \multirow{5}{4.1cm}{{Aim also at maximizing throughput.}} \\  
\cline{2-4}	
 	&   \citet{bib:NOVA} & X & & \\ 
\cline{2-4}	
	&   \citet{bib:oyman} &  X & &  \\ 
\cline{2-4}	
	&   \citet{bib:multicast} &  X & &  \\ 
\cline{2-4}	
	&   \citet{bib:buffered_video} &  X & Playout stalls &  \\
\cline{2-5}	
	&   \citet{bib:fred} & X & & \multirow{1}{4.1cm}{{Tries to maximize buffer filling.}} \\ 
\cline{2-5}	
	&   \citet{bib:weighted} & X & & \multirow{5}{4.1cm}{{Proxy-based solutions.}} \\ 
\cline{2-4}	
	&   \citet{bib:essaili} &  X & &  \\ 
\cline{2-4}	
	&   \citet{bib:dash_friendly} &  X & &  \\ 
\cline{2-4}	
	&   \citet{bib:tiantian} &  X & &  \\ 
\cline{2-4}	
	&   \citet{bib:jnca_extra4} &  X & &  \\ 
\hline

\multirow{1}{*}{{Web browsing}} 	&   \citet{bib:web_2016} &    & Delay \& Page state info. & \multirow{1}{4.1cm}{{  }} \\ 
\hline

\multirow{1}{*}{{Multi-service}} 	&   \citet{bib:renault} &    & Delay, Packet loss rate \& Jitter  & \multirow{1}{4.1cm}{{  }} \\ 
\hline

\end{tabular}
\label{tab:passive_user}
\end{table*}

Some authors also presented scheduling techniques that follow a methodology that is different from what was seen so far; more precisely, the algorithms proposed in \citep{bib:weighted,bib:essaili,bib:dash_friendly,bib:tiantian,bib:jnca_extra4} adopt a proxy-based approach, in which not only users with low buffer levels are prioritized, but also the proxy is able to modify the request of a client and serve a video segment with a lower bitrate, namely if this last procedure is able to avoid impending playout stalls. 
In spite of the fact that the use of this type of proxy is very helpful in maximizing considerably the number of interruption-free transmissions, there are also some issues that arise and that must be considered before implementing it: first, the video streaming system needs to provide a seamless  switching between different bitrate representations, of the same multimedia content, during a streaming session (e.g., a DASH-based approach could be followed); secondly, a client may not be willing to accept a bitrate representation which is different from the one it requested (for instance, although the DASH specifications foresee the case where alternative representations could be admissible at the client side, this option can only be used if the client activates it first).

\subsubsection{Other Applications}

Turning the attention to applications other than video streaming, a wireless resources redistribution algorithm is proposed in \citep{bib:web_2016} for web browsing, where the page state information is adopted as one of the inputs of the scheduler --- more specifically, the proposed method subtracts some capacity from a client when the respective web page enters the ``interactive'' state, i.e., when something first renders on the screen and the respective user can start browsing; afterwards, the liberated resources are redistributed to those clients that have web pages in the ``loading'' state, i.e., the initial downloading part where the respective users are awaiting something to be displayed and which corresponds to the most sensitive period in terms of QoE.
\citet{bib:renault} presented a scheduling strategy for real-time services (especially for voice, but it can also be used for video flows), which follows a PF-like approach with the addition of a multiplying term that takes into account a QoE metric computed at the devices and reported to the scheduler (this QoE metric depends on the delay, packet loss rate and network jitter); thus, the users that have higher a QoE have a higher probability of being prioritized under this scheduling algorithm, which could pose some difficulties in terms of fairness.

\subsubsection{General Considerations on Active End-user Device / Passive User Strategies}

As can be inferred from the QoE-oriented scheduling solutions overview performed in this subsection, which are summarized in Table~\ref{tab:passive_user}, the possibility of reporting relevant QoE metrics, from the users' devices to a scheduler, is mainly useful for video streaming services. More specifically, the report of information that is only available at the device, such as the buffer level, enables to design scheduling algorithms that are more capable of attaining the interruption-free video goal, thus providing an enhanced QoE when compared to passive end-user device techniques.
Moreover, it is noteworthy to point out the usefulness of using real QoE metrics instead of estimated ones: for instance, even though a scheduler could try to estimate the users' buffer level at the network side (i.e., without any buffer level report from the end-user devices to the network, which would avoid the need for feedback channels), several factors may sometimes turn this estimation task into a meaningless one --- e.g., different initial playout delays (an information which is usually not known by the network) lead to different estimated buffer levels, as well as when a user pauses a video but the respective video download is not interrupted.

\begin{table*} [!tb] 
\footnotesize
\renewcommand{\arraystretch}{1.3}
\centering
\caption{Active End-user Device / Active User Strategies.}
\vspace{8pt}

\begin{tabular}{| c || c || c |} 
\hline
\multirow{2}{*}{\textbf{Application}} & \multirow{2}{*}{\textbf{Reference}} & \multirow{2}{*}{\textbf{QoE optimization based on}} \\
& & \\
\hline\hline

\multirow{2}{*}{{Multi-service}} 	&  \citet{bib:12_of_001} & Users' quality preference indication \\ 
\cline{2-3}	
	&   \citet{bib:one_bit_feedback} & Users' satisfaction feedback \\ 
\hline

\end{tabular}
\label{tab:active_user}
\end{table*}

On the other hand, the reported QoE metrics should be incorporated carefully by the different scheduling methods, i.e., these should take into account the different states of each service and should not regard the same information always in the same way, so that the final QoE is optimized --- for instance, consider a scheduler that prioritizes the users that have a low buffered playout time; noticing that a playout stall is usually more annoying than a slightly longer initial playout delay, this scheduler should refrain from giving a high priority to those users that are starting a video download (although their buffer level is closer to zero), because otherwise this could entail a lack of resources for already existing flows.


\subsection{Active End-user Device / Active User Strategies}
\label{subsection:active_user}  

As mentioned in the Introduction, humans have the decisive judgment about the received service quality. Therefore, scheduling strategies that exploit direct and conscious inputs of the users have the advantage of knowing their preferences and if they are satisfied with the service. On the other hand, these approaches require that end-user devices are capable of receiving the necessary user inputs. The strategies described next (and summarized in Table~\ref{tab:active_user}) adopt this active user approach.

\citet{bib:12_of_001} presented a QoE provisioning method that enables users to indicate their preference in a dynamic and asynchronous manner concerning the instantaneous perception of the service performance. To attain this goal, a Graphical User Interface (GUI) displays and captures the users' options (increase or reduce quality), the feasibility and the repercussions of their act (cost). When the preference of a user is manifested, the features of the utility functions are dynamically adjusted, in order to exploit the NUM theory by enabling the smooth incorporation of users' subjective decision in the resource allocation process. 
A QoE-aware scheduling framework is proposed in \citep{bib:one_bit_feedback} to maximize the average amount of satisfied users, where it is assumed that users can feedback one bit to express their degree of satisfaction. Since the decisions taken by the users have a direct influence on the QoE enhancement process, non-trivial fairness constraints are also added so as to prevent starvation.


\subsection{Other QoE-oriented Scheduling Methods}
\label{subsection:other_sched}

All the previous scheduling algorithms addressed the downlink scenario, with the goal of maximizing the aggregate experienced quality of all users within one cell of a single operator. Nonetheless, QoE-aware scheduling methods can enhance the wireless resources management in other scenarios, such as the uplink direction, the multi-cell case, under heterogeneous, cognitive radio, relay and MIMO networks, as well as when dealing with energy-related issues.

\subsubsection{Uplink}

A resource scheduling framework for live video uplinking is proposed in \citep{bib:uplink}, in which more resources are allocated for popular contents while ensuring a certain QoE level (based on throughput) for the less popular ones. 
\citet{bib:surveillance} presented a QoE-driven scheduler for uplink unicast video delivery (intended for surveillance systems), in which the required throughput and delay are the main factors taken into account by the resource allocation scheme. 
In \citep{bib:uplink_carrier_aggreg}, a QoE-based joint resource allocation method is proposed for uplink scenario of LTE networks that make use of carrier aggregation, where the assessment of the users' QoE satisfaction degree is performed by a resource cost, link reward and utility function designed by the authors.
The works presented in \citep{bib:teleoperation, bib:tele_haptic} address haptic teleoperation over wireless networks and introduce new allocations algorithms regarding the scheduling process in the uplink direction, which aim at improving the QoE by reducing the communication delay.
With respect to disaster scenarios, and assuming that base stations are mounted on Unmanned Aerial Vehicles (UAVs), an uplink scheduling procedure is proposed in \citep{bib:disaster} that has the goal of enhancing the QoE   of critical users (i.e., those that are in danger positions or have low battery power), so that they can better communicate with the outside world.

\subsubsection{Multi-cell}

The scheduling strategies mentioned up to this point consider the situation where each base station serves as an independent scheduler. However, a centralized network controller can be adopted in order to improve the QoE of those users that can connect to more than one base station.
The works presented in \citep{bib:multicell_2,bib:too_math} addressed the inter-cell interference problem and devised QoE-oriented resource allocation techniques that aim at enhancing the users' satisfaction and fairness, especially for cell edge users.
\citet{bib:3in1} proposed a QoE-aware scheduling algorithm that also incorporates admission and handover procedures (to neighboring base stations), in order to try to ensure that video streaming users that connected previously to the network maintain at least an acceptable QoE (namely by taking into account the buffer level).
In \citep{bib:control_approach}, the resource allocation problem is considered for a high number of simultaneous video streaming sessions within a dense wireless network scenario, in which a central scheduler has the goal of providing the allocation configuration (among all base stations) that enables to improve the users' QoE.

\subsubsection{Heterogeneous, Cognitive Radio, Relay \& Multi-user MIMO Networks}

\begin{figure}[!tb]
\begin{center}
\includegraphics[ width=0.566\columnwidth , trim= 0 0 0 0 ]{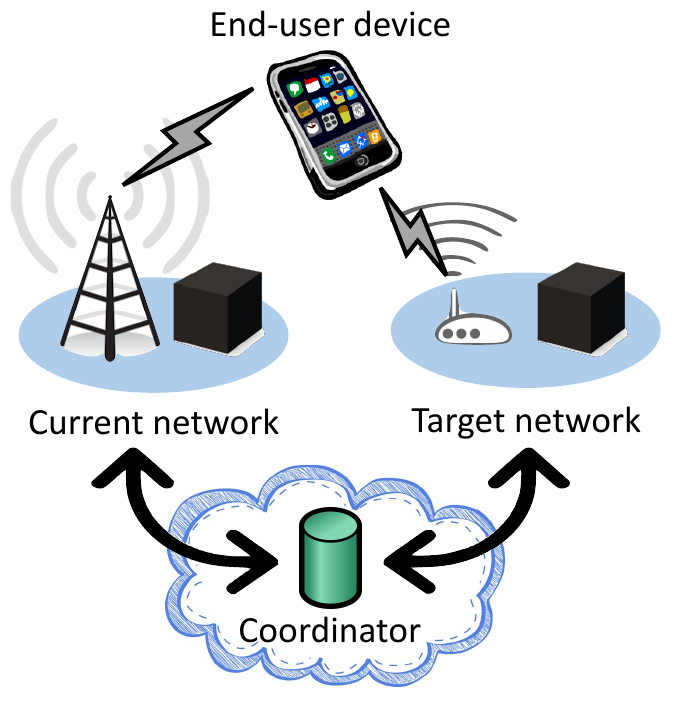}
\end{center}
\caption{General architecture for heterogeneous network management.}
\label{fig:hetnet}
\end{figure}

Only single operators have been considered so far, but some research has already been done regarding heterogeneous wireless networks, where users have the desire of being connected to the best accessible networks according to each user application specifications and personal preferences. 
For instance, \citet{bib:6_of_15_of_001} adopted a game-theoretic approach as a framework for user satisfaction based wireless resource scheduling with various service providers, different sorts of users and several service categories. 
In \citep{bib:handover}, a QoE-based handover architecture is presented for heterogeneous mobile networks, which has the goal of providing seamless mobility in multi-operator and multi-access systems while ensuring that users have the best possible connection in terms QoE.
Regarding mobile traffic offloading, a method to improve the QoE of video delivery, which incorporates a collaboration between Wi-Fi hotspots and LTE base stations, is proposed in \citep{bib:offloading}, where the subjective quality assessment is based on video playback discontinuities.
\citet{bib:hetnet} devised a QoE-based scheduling algorithm that aims at enhancing the video delivery over heterogeneous mobile networks, namely by limiting the inter-cell interference experienced by some users.
The work presented in \citep{bib:jnca_extra3} addresses the real-time traffic splitting across cellular and Wi-Fi heterogeneous networks (focusing on video streaming applications) and provides a solution for resource allocation that enhances QoE while reducing delay and energy consumption (namely of mobile terminals). 
Regardless of each specific scheduling solution for heterogeneous networks, all of them have a major difference when compared to the traditional scheduling methods: as depicted in Fig.~\ref{fig:hetnet}, an extra entity is required --- indicated in the figure as ``Coordinator'' --- as well as extra signaling is needed, in order to ensure a proper QoE-oriented scheduling in all instances, namely when performing inter-network handover or traffic splitting. Since different networks can make use of different technologies, or can even be associated with different service providers, it is also not possible, within heterogeneous networks, to make use of a single scheduler and apply the traditional scheduling solutions. Hence, not only each network has its own scheduler, but also the ``Coordinator'' has the complex task of coordinating these schedulers (which usually includes conveying relevant QoE information among them), in order to enhance the users' QoE or maintain at least an acceptable QoE.

Device-to-Device (D2D) systems can also benefit from QoE-aware approaches.
In a somewhat similar fashion to the heterogeneous case, D2D networks also require a ``Coordinator'' unit (as well as extra signaling) for an effective QoE-oriented scheduling. More specifically, not only this unit is responsible for collecting information sent wirelessly from the devices (thus the extra signaling is less reliable than the wired one of heterogeneous networks), but it also performs important scheduling tasks, such as which and when D2D pairs can access the channel or which data should be conveyed by each device.
In \citep{bib:D2D}, a QoE-driven allocation technique is proposed for video streaming through D2D transmissions, where the wireless resource scheduling goal is to enhance the time-averaged quality of video streams transmitted over D2D communications, while constraining the number of stall events for each stream.
\citet{bib:D2D_fog} devised a QoE-aware D2D-based mobile task outsourcing, in which the cellular mobile equipments cooperate so as to perform computational intensive tasks; a special node within the cellular network is responsible for the D2D task scheduling, where the maximum wait time of each task is adopted as the main QoE influence factor.
A management procedure is presented in \citep{bib:jnca_extra5} for mode switching between base station approach and D2D communications, where QoE parameters are the basis for the switching decisions, as well as for the inclusion, or not, of a third-party terminal to assist a D2D transmission.
In order to provide greater flexibility within D2D systems,
\citet{bib:extra_2019_a} proposed the categorization of D2D users into different groups (based on their practical application or service), 	which enables to define a suitable utility function for each group, followed by a solution for QoE-oriented resource allocation, which is formulated as a dynamic Stackelberg game that considers multi-criteria decision making.
With respect to vehicular networks, some works provide QoE-driven frameworks, namely for video-on-demand over urban vehicular networks \citep{bib:vanet_0}, for scalable video streaming over cooperative Vehicle-to-Vehicle (V2V) and Vehicle-to-Infrastructure (V2I) communications \citep{bib:vanet_svc}, as well as with the goal of reducing the transmission delay of vehicular security applications \citep{bib:vanet_delay}.

\begin{figure}[!tb]
\begin{center}
\includegraphics[ width=0.65\columnwidth , trim= 0 0 0 0 ]{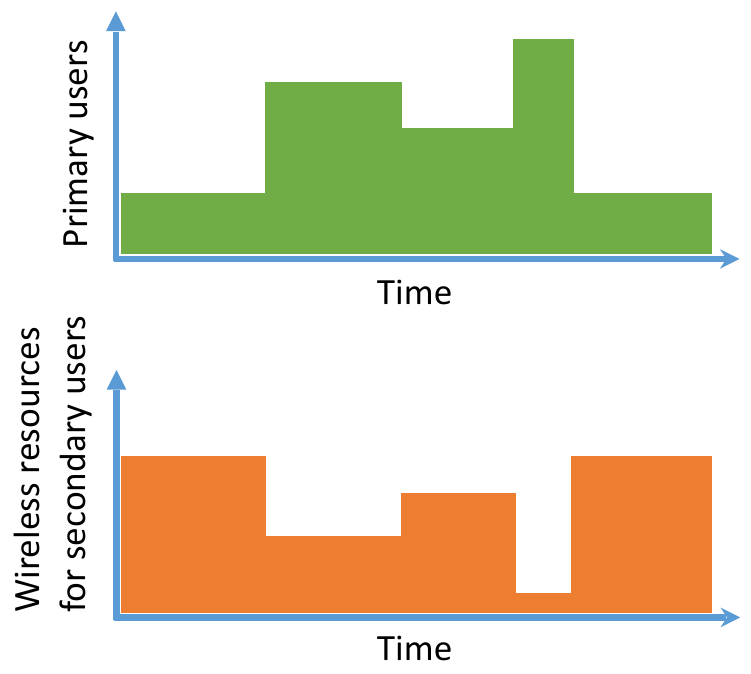}
\end{center}
\caption{Dynamic availability of wireless resources within cognitive radio networks.}
\label{fig:cog_resources}
\end{figure}

QoE provisioning schemes can also be used when addressing other wireless resources scheduling problems, such as in cognitive radio networks. 
It is noteworthy to mention that, on the one hand, the opportunistic spectrum access of cognitive radio approaches can lead to an increase of the total available wireless resources that can be scheduled for the secondary users, but, on the other hand, the resources allocated to the secondary users can be re-occupied by the primary users at any time. Hence, not only this leads to an unstable availability of wireless resources for the secondary users (as depicted in Fig.~\ref{fig:cog_resources}), but also this dynamic behavior poses an additional challenge to the wireless system scheduler.
\citet{bib:cog_radio_2} developed a channel assigning algorithm for the transmission of multimedia content over cognitive radio systems, where the base station allocates available channels to secondary users according to the respective QoE requirements (delay and multimedia content quality). 
A jointly design of spectrum sensing and access policies for multi-user QoE-oriented video delivery within cognitive radio networks is presented in \citep{bib:cog_2016}, in which the goal is to achieve fairness among the users while maximizing the average QoE (which is based on throughput). 
\citet{bib:potencial_game} addressed networks where different types of base stations are deployed to exploit a heterogeneous spectrum pool, containing licensed and harvested spectrum, in order to propose a game-theoretic approach that solves the problem of optimizing the global users' satisfaction (based on their throughput requirement) by jointly optimizing spectrum sharing, user scheduling, and power allocation in a decentralized manner.
A framework is devised in \citep{bib:cog_5g} that has the goal of managing the inevitable spectrum handoff within cognitive networks while providing seamless multimedia content streaming and QoE enhancement (namely by taking into consideration the delay and the quality of the multimedia content).
In \citep{bib:lin_5g}, a QoE-oriented dynamic channel access procedure is presented in order to handle the sharing of spectrum between licensed and secondary users; the scheduler takes into account the type of service that is being requested by the secondary users (either delay sensitive or not) and aims at minimizing the average queuing time of the respective packets.
\citet{bib:sug_1} proposed a QoE-driven rate control and resource allocation scheme for cognitive Machine-to-Machine (M2M) communication; the authors focused on how to maximize the QoE of all M2M pairs, namely by employing a stochastic optimization model in which the user-perceived application quality metric is associated with data rates.

With respect to wireless relaying in cognitive radio networks, and by considering QoE indicators (relay buffer status) rather than traditional QoS indicators, \citet{bib:cog_radio} showed that better user experience can be achieved with sub-optimum system capacity. 
\citet{bib:mesh} considered multi-hop wireless networks and introduced a scheduling algorithm that jointly performs an optimization of multiple video, audio and data streams transmission based on certain QoE targets. 
The works presented in \citep{bib:relay_fast_video,bib:ad_hoc} developed cross-layer resource allocation schemes for two-hop and ad hoc networks, respectively, where the goal of both works is to enhance the QoE by minimizing the end-to-end video delivery time. 
In \citep{bib:mu_mimo}, an efficient system is proposed for video streaming over a wireless system composed by a large amount of wireless helper equipments with multi-user MIMO capabilities, where QoE metrics like video quality and rebuffering percentage are used to optimize the transmission scheduling of users at each base station. 
Other examples of QoE-based scheduling for multi-user MIMO systems are given in \citep{bib:zfbf_mimo, bib:mimo_chino}, where user selection procedures based on transmitted rate and delay are proposed in order to maximize the average multi-service satisfaction degree.
With respect to Virtual Reality (VR) over multi-user MIMO networks, \citet{bib:vr} devised a resource allocation technique based on a maximum aggregate delay-capacity utility function, which aims at reducing the delay between user motion and the presentation of multimedia content, as well it aims at maximizing the number of connected VR users with an acceptable QoE.

\subsubsection{Energy-related Issues}

The ``green'' networks is also a domain where QoE-based wireless resources management can have an important role. 
For instance, \citet{bib:distortion} proposed a framework to optimize the power allocation at the base station side, in which the adopted utility function --- formulated as
\begin{align}
 U_i\left( \sum_n p_{i,n}\right) = \frac{MOS_i}{\sum\limits_n p_{i,n}},
\end{align}
where $p_{i,n}$ stands for the allocated power to the $i^\text{th}$ user on the $n^\text{th}$ sub-channel and $MOS_i$ denotes the MOS of the  $i^\text{th}$ user --- aims at providing a balance between energy efficiency and QoE for video streaming users.
\citet{bib:last_5} devised a power allocation optimization method regarding the QoE for multi-services, which is intended to solve the following problem (where the goal is to minimize the overall power consumption, while ensuring a good level of perceived quality to each user):
\begin{align}
\begin{split}
	{\text{minimize}} &\quad P_\text{total}=\sum\limits_{i}\sum\limits_{n} p_{i,n} \\
	\text{subject to} &\quad \Psi_i \left(\sum\limits_{n} p_{i,n} \right) \geq MOS_{i\_\min} \\
	&\quad P_\text{total} \leq P_{\max} \\
	&\quad p_{i,n} \geq 0, \forall_{i,n} 
\end{split},
\end{align}
where $P_\text{total}$ represents the total power allocated to the users, $P_{\max}$ corresponds to the system maximum power, 
$\Psi_i(\cdot)$ denotes the mapping relationship between $p_{i,n}$ and a MOS value regarding the multimedia service chosen by the $i^\text{th}$ user, and $MOS_{i\_\min}$ represents the minimum acceptable MOS also concerning the multimedia service chosen by the $i^\text{th}$ user.
In \citep{bib:renewable}, an ``energy-source'' based method for delivering multimedia content asynchronously is presented, which adjusts the consignment of delay-tolerant content according to the time intervals where the renewable energy is available, i.e., it tries to balance the service provider energy costs and the QoE (namely the delay) experienced by users.
An approach to power-cycle base stations and to control the playback of video streams is presented in \citep{bib:power_bs}, where the goal is to reduce the overall energy consumed by the base stations while maintaining a high QoE for the users.
A framework to reduce small cell energy consumption contingent on QoE restrictions has been proposed in \citep{bib:save_energy}, where an analytical investigation is performed regarding the trade-off between switching off underloaded base stations and the performance degradation experienced by the users.  
An energy efficient resource on-off switching framework for a cellular network comprising a femtocell at the cell edge of a macrocell is investigated in \citep{bib:energy_hete}, where the goal is to minimize the energy consumption of the cellular network while satisfying a desired level of QoE, which is defined as buffer starvation probability of a mobile device.
In \citet{bib:extra_2019_f}, an energy-efficient scheduling technique is devised in order to decrease energy consumption of wireless systems, such as heterogeneous networks and dense femtocell networks, as well as to mitigate the cell edge interference, in which a QoE-driven strategy considers resources demand at every pool (instead of solely depending on the signal-to-noise ratio) before shifting a user from one pool to another.
\citet{bib:extra_2019_b} proposed a resource and re-association scheduling method based on Benders' decomposition to decrease the energy consumption within Wireless Local Area Networks (WLANs), namely by aggregating users to fewer Access Points (APs) and by turning off many APs without compromising the users' QoE as well as the network coverage.

With respect to terminals energy consumption, in \citep{bib:sleep_video}, a framework is presented that decreases the energy used by mobile equipments during video delivery over WLANs, in which a QoE-based algorithm makes an estimation of the video quality perceived by a user and adapts the sleep periods of the wireless device so as to enhance the power efficiency while preserving video quality. 
\citet{bib:energy_VoWLAN} proposed a novel approach for energy conservation in mobile terminals regarding VoIP flows over WLANs, namely by tuning the sleep intervals according to the user QoE.
Schedulers that make use of the discontinuous reception method for LTE systems have been proposed in \citep{bib:drx, bib:jnca_extra2} (with the latter focusing on VoIP applications), which can provide battery saving for the terminals without degrading the QoE of the services, namely the delay perceived by a user.
In \citep{bib:mobile_offloading}, a scheduling policy is proposed to save energy of mobile devices via optimal transmission scheduling of mobile-to-cloud task offloading, which considers several QoE domains in order to capture the energy-latency-pricing trade-off.
\citet{bib:extra_2019_e} also addressed computation offloading but with respect to UAV cloud system, namely by devising a resource allocation algorithm with heterogeneous QoE support which is able to enhance the energy efficiency of the UAVs.
As previously mentioned, \citet{bib:jnca_extra3} proposed a method that reduces energy consumption of mobile terminals regarding cellular and Wi-Fi heterogeneous networks.

\begin{table*} [!tb] 
\footnotesize
\renewcommand{\arraystretch}{1.3}
\centering
\caption{Other QoE-oriented Scheduling Methods.}
\vspace{8pt}

\begin{tabular}{| c | c |}
\hline
\multirow{2}{*}{\textbf{Scope}} & \multirow{2}{*}{\textbf{References}} \\
&  \\

\hline\hline
\multirow{2}{*}{Uplink} & \multirow{2}{13.000cm}{\citet{bib:uplink, bib:surveillance, bib:uplink_carrier_aggreg, bib:teleoperation, bib:tele_haptic, bib:disaster} } \\
&  \\
\hline
\multirow{2}{*}{Multi-cell} & \multirow{2}{13.000cm}{\citet{bib:multicell_2, bib:too_math, bib:3in1, bib:control_approach} } \\
&  \\
\hline
\multirow{2}{*}{\makecell{Heterogeneous \\ networks}} & \multirow{2}{13.000cm}{\citet{bib:6_of_15_of_001, bib:handover, bib:offloading, bib:hetnet, bib:jnca_extra3} } \\
 &  \\
\hline
\multirow{2}{*}{\makecell{Device-to-device\\communications}} & \multirow{2}{13.000cm}{\citet{bib:D2D, bib:D2D_fog,bib:jnca_extra5, bib:extra_2019_a} } \\
 &  \\
\hline
\multirow{2}{*}{Vehicular networks} & \multirow{2}{13.000cm}{\citet{bib:vanet_0, bib:vanet_svc, bib:vanet_delay} } \\
&  \\
\hline
\multirow{2}{*}{\makecell{Cognitive radio\\networks}} & \multirow{2}{13.000cm}{\citet{bib:cog_radio_2, bib:cog_radio, bib:cog_2016, bib:potencial_game, bib:cog_5g, bib:lin_5g, bib:sug_1} } \\
 &  \\
\hline
\multirow{2}{*}{Relay networks} & \multirow{2}{13.000cm}{\citet{bib:mesh, bib:cog_radio, bib:mu_mimo, bib:relay_fast_video, bib:ad_hoc} } \\
&  \\
\hline
\multirow{2}{*}{\makecell{Multi-user MIMO\\networks}} & \multirow{2}{13.000cm}{\citet{ bib:zfbf_mimo, bib:mu_mimo, bib:mimo_chino, bib:vr} } \\
 &  \\
\hline
\multirow{2}{*}{\makecell{Base stations energy\\consumption}} & \multirow{2}{13.000cm}{\citet{bib:distortion, bib:last_5, bib:renewable, bib:power_bs, bib:save_energy, bib:energy_hete, bib:extra_2019_f, bib:extra_2019_b} } \\
 &  \\
\hline
\multirow{2}{*}{\makecell{Terminals energy\\consumption}} & \multirow{2}{13.000cm}{\citet{bib:sleep_video, bib:energy_VoWLAN, bib:drx, bib:jnca_extra2, bib:jnca_extra3, bib:mobile_offloading, bib:extra_2019_e} } \\ 
 &  \\
\hline
\end{tabular}
\label{tab:other_sched}
\end{table*}

\subsubsection{General Considerations on Other QoE-oriented Scheduling Methods}

The scheduling strategies mentioned in this subsection, which are also summarized in Table~\ref{tab:other_sched}, clearly show that QoE-aware decisions are useful in a vast field of resource management regarding wireless communications. 
On the one hand, these scenarios (other than the downlink direction with respect to a single base station) introduce some particularities that must be taken into account if one wants to adopt the techniques that were presented in Sections~\ref{subsection:passive_device}, \ref{subsection:passive_user} and \ref{subsection:active_user} --- issues like, e.g., the exchange of information between different wireless technologies when heterogeneous networks are considered, or the dynamic nature of cognitive radio networks, among others, can lead to a more challenging scheduling scenario.
On the other hand, the same particular characteristics can also provide new tools that enable to leverage the users' QoE --- for instance, the handover procedure that is available for the centralized approach of the multi-cell scenario (namely for those users that can be served by more than one base station simultaneously) can be activated with the purpose of liberating resources from a base station, thus enabling that additional users are able to achieve a higher QoE.

Still within the scope of other QoE-oriented scheduling methods, it is noteworthy to mention that the QoE might also be improved if some scheduling decisions are conveyed to the users, such as by giving the option of saving energy at the terminal side at the cost of, e.g., a slightly inferior video streaming quality.


\section{Discussion: Challenges, Issues \& Future Directions}
\label{section:disc_future}  

The scheduling methods referred above perform resource allocation and/or prioritization at the wireless MAC/physical layer level, which is the scope of this work. Nevertheless, there are also QoE-aware algorithms that, although taking into consideration the wireless channel specificities, they permit to preserve an entirely standard and application-layer unaware physical/MAC operation --- for instance, algorithms that run above the MAC layer and which arrange suitably the order of the data units delivered to this layer; more examples and details can be found in \citep{bib:unaware_1,bib:unaware_4,bib:unaware_2,bib:unaware_5,bib:unaware_2016b,bib:unaware_2016a,bib:vulkan,bib:kumar_sur,bib:unaware_2016d,bib:unaware_2016c,bib:ratio_req,bib:extra_2019_d,bib:natal_1}.

\begin{figure}[!tb]
\begin{center}
\includegraphics[ width=0.9\columnwidth , trim= 0 0 0 0 ]{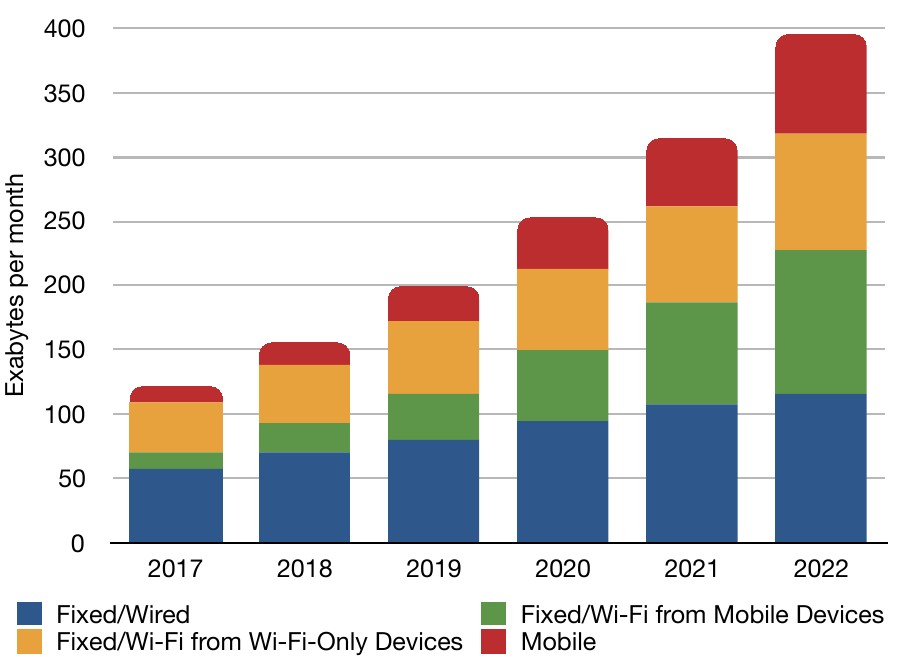}
\end{center}
\caption{Global data traffic, wired and wireless \citep{bib:cisco}.}
\label{fig:traffic_growth}
\end{figure}

Taking into account the scheduling strategies surveyed herein, video streaming is clearly the application that can benefit more from QoE-aware schedulers. In addition, this will keep being an important topic of research, since not only video streaming is now the most popular application on the Internet, having been responsible for 59\% of the entire mobile data traffic at the end of 2017, but also it is foreseen that over three-fourths (79\%) of the worldwide mobile data traffic will be video by 2022 \citep{bib:cisco}. 
Moreover, as depicted in Fig.~\ref{fig:traffic_growth}, global mobile data traffic will grow 7-fold from 2017 to 2022, a compound annual growth rate of 46\%, thus reaching 77.5 exabytes per month by 2022, up from 11.5 exabytes per month in 2017, which means that mobile data traffic will grow 2.0 times faster than fixed traffic from 2017 to 2022.
On top of that, some authors already suggested that multimedia services should be priced based on the QoE rather than based on the transmitted binary data \citep{bib:price_qoe}.

Nevertheless, it was also seen that a scheduling algorithm and its underlying QoE estimation model need to be careful adjusted for each application, since the user's perception regarding different applications is influenced by different factors. In other words, a ``one-fits-all'' solution for wireless resource schedulers is not feasible when QoE comes into action, as quality expectations are highly service dependent, which means that scheduling strategies should always take into account the intended application. As a rule of thumb, schedulers should try to avoid buffer emptiness when dealing with video streaming, they should reduce the delay as much as possible for VoIP and web browsing services, whereas if they are unaware of the application, then serving the best possible throughput for each user is the most reasonable option.

Another issue that follows the previous considerations is that the performance of different QoE-oriented schedulers might not be easily comparable, mainly due to the fact that a common reference scenario might be impractical when dealing with different applications. Accordingly, it is important to clearly identify which are the specific goals of each solution, in order to determine which ones can be adopted for the same scenario; otherwise, the performance comparison becomes unfair, even for applications that appear to be similar --- for instance, and considering the video streaming application, a QoE-based scheduling algorithm that was expressly adjusted to handle the transmission of certain videos, namely those that are stored in a repository, might underperform in other video streaming scenarios, e.g., live transmissions. 

It is noteworthy to point out that admission control was not taken into account by almost all the scheduling algorithms surveyed herein (the work of \citet{bib:3in1} is the only exception). Admission control can play an important role in terms of QoE provisioning, namely by admitting a new user only if the already connected users and the new user can have an acceptable QoE. Although it seems advantageous to jointly design QoE-oriented scheduling strategies and admission control mechanisms, the vast majority of the scheduling methods referred above only studied the impact of the number of users on the QoE; nevertheless, these studies could be used to infer the number of users that can be accommodated in the wireless network, thus serving as an input when devising admission control procedures. On the other hand, only a few authors have presented admission control solutions that are QoE-aware --- cf. \citep{bib:ad_1,bib:ad_2,bib:ad_3,bib:ad_4,bib:ad_5}. One possible explanation for the lack of literature about this topic is the fact that, as found by \citet{bib:ad_comp} (which evaluated and compared QoE-based admission control mechanisms), there is an inherent difficulty to precisely calibrate the algorithms in order to obtain satisfying QoE-oriented admission decisions, namely because not only the best calibration, but also the overall performance of the admission control algorithms varies strongly with the underlying scenario. 


With respect to the three categories of QoE-aware schedulers addressed in this work, there is still room for further developments in all of them, but the active end-user device / passive user strategies are the ones which might be the focus of attention in the future. The main reason for this is that they enable to gather the maximum amount of relevant QoE information as close as possible to the end-user without requiring direct and conscious human QoE inputs; accordingly, they are the clear candidates to better enhance the performance of QoE-oriented schedulers without annoying the end-user with QoE assessment prompts. 
Moreover, with the support of miscellaneous biometric sensors that are being incorporated in smartphones --- e.g., the TrueDepth camera system of iPhone~X \citep{bib:iphone}, which is able to analyze
more than 50 different facial muscle movements, including attention confirmation by detecting the direction of the user's gaze --- more accurate QoE data are becoming available at the end-user device that stem from live human biometric indirect inputs --- like pupil variations and galvanic skin reactions \citep{bib:pupilas}, facial expressions \citep{bib:facial_express}, and body gesticulations \citep{bib:body_gestures} --- which can provide a viable online QoE optimization scheme in a model-free manner \citep{bib:sug_3}.
Nonetheless, and in spite of the fact that there are already several feedback mechanisms that allow to report a significant number of parameters from the end-user devices to the scheduler, as in the 3GP-DASH standard, important challenges still arise: which are the relevant QoE parameters and which is the specific relationship between each one of them and the user's subjective perception? Within this subject, machine learning methods can play a major role, higher than the one seen so far, as they enable to devise complex models and learn about non-obvious correlations between the collected parameters and the users' QoE. In addition, these computer science techniques are a powerful tool to better deal with particular issues regarding mobile networks, such as phone battery, phone overheat, or data connectivity costs, which make QoE assessment more demanding when compared to fixed networks.

\begin{figure}[!tb]
\begin{center}
\includegraphics[ width=0.9\columnwidth , trim= 0 0 0 0 ]{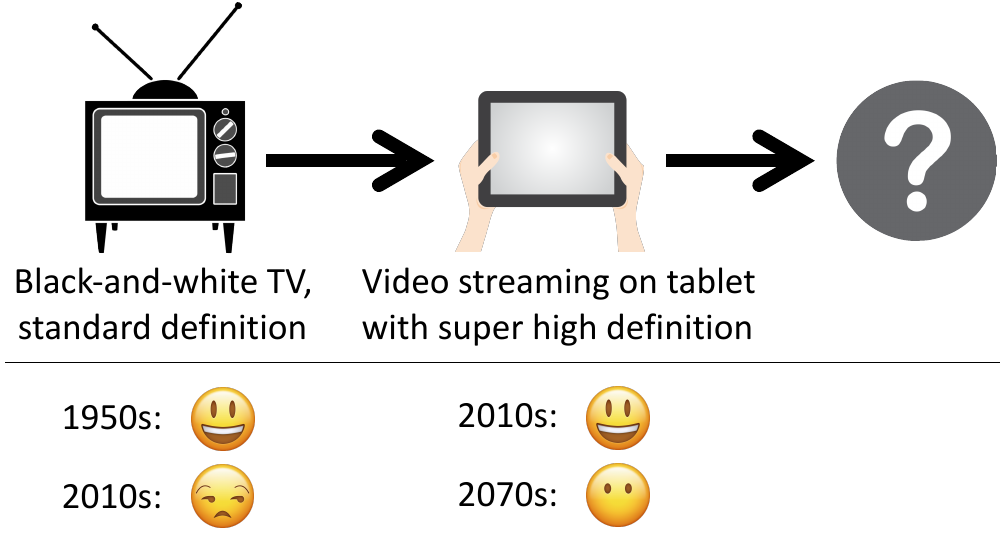}
\end{center}
\caption{Example of QoE evolution over time according to available technology.}
\label{fig:qoe_evolution}
\end{figure}

Last but not least, QoE evolves over time in the same manner as technology itself does: for instance, and as illustrated in Fig.~\ref{fig:qoe_evolution}, although today a user may require super high definition video content in order to say that he is experiencing a great QoE, some decades ago the same subjective perceived quality would be true for a TV transmission with standard definition. Consequently, a scheduler that takes into account a QoE model that reflects today's reality may not be the most appropriate one in the future, meaning that QoE-oriented schedulers will always attract the research community's attention.

\section{Conclusions}
\label{section:conclusions}   

Wireless resources scheduling is starting to become more user-centric QoE-oriented, replacing the traditional system-centric QoS-driven approach. Thus, the network operators, in addition to multimedia service providers, aspire to have faithful models that are able to evaluate, foresee and even manage QoE, specially for multimedia content transmissions. This paper provided an extensive survey about this research topic: first, QoE was explained, namely the factors that influence the user experience, as well as some QoE estimation methods regarding multimedia services over communication systems; next, the evolution of wireless scheduling techniques was presented, including QoS-QoE mapping strategies and utility-based optimization; finally, state-of-the-art QoE-based scheduling strategies for wireless systems were described, highlighting the application/service of each solution, as well as the parameters adopted for QoE optimization. Since there are still many issues to be explored and resolved, it is foreseen that a lot of research activity will surely be performed in the future, in order to expand the research frontier of wireless resources schedulers.




  
\section*{Acknowledgments}
This work was funded by Instituto de Telecomunicações and Fundação para a Ciência e a Tecnologia (FCT) under projects UID/EEA/50008/2019 and OCTHOPUS, and by project MESMOQoE (no. 023110 -- 16/SI/2016) supported by Norte Portugal Regional Operational Programme (NORTE 2020), under the PORTUGAL 2020 Partnership Agreement, through the European Regional Development Fund (ERDF).


\bibliographystyle{elsarticle-harv}
\bibliography{bilbos} 


\end{document}